\RequirePackage{fix-cm}
\documentclass[a4, twocolumn, epjc3, comma, sort&compress, natbib]{svjour3}
\bibpunct{[}{]}{,}{n}{}{,} 

\journalname{Eur. Phys. J. }

\usepackage{latexsym}
\usepackage{amsmath}
\usepackage{amssymb}
\usepackage{amsfonts}
\usepackage{CJKutf8}
\usepackage{dsfont}
\usepackage{enumitem}
\usepackage{slashed}
\usepackage{bm}
\usepackage{wasysym}

\usepackage[mathscr,scaled=1.15]{urwchancal}
\DeclareFontFamily{OT1}{pzc}{}
\DeclareFontShape{OT1}{pzc}{m}{it}%
{<-> s * [1.15] pzcmi7t}{}
\DeclareMathAlphabet{\mathpzc}{OT1}{pzc}{m}{it}

\usepackage{color}
\definecolor{purple}{rgb}{0.5,0,0.5}
\definecolor{blue}{rgb}{0.0,0,0.9}
\definecolor{prdblue}{rgb}{0.133,0.118,0.498}
\usepackage[colorlinks=true, pdfstartview=FitV, linkcolor=prdblue, citecolor= prdblue, urlcolor=prdblue]{hyperref}

\usepackage{enumitem}
\usepackage{multirow}
\usepackage{subfigure}
\usepackage{textcomp}

\usepackage{supertabular}
\usepackage{placeins}
\usepackage{epsfig}
\usepackage{graphicx}

\usepackage{dcolumn,booktabs}
\newcolumntype{d}[1]{D{.}{.}{#1}}

\begin{document}
\begin{CJK}{UTF8}{song}

\title{$\,$\\[-6ex]\hspace*{\fill}{\normalsize{\sf\emph{Preprint no}.\ NJU-INP 060/22}}\\[1ex]
Heavy+heavy and heavy+light pseudoscalar to vector semileptonic transitions}

\author{Hui-Yu Xing\thanksref{NJU,INP,eHYX}%
       $\,^{\href{https://orcid.org/0000-0002-0719-7526}{\textcolor[rgb]{0.00,1.00,0.00}{\sf ID}}}$
        \and
       Zhen-Ni Xu\thanksref{NJU,INP,eZNX}%
       $\,^{\href{https://orcid.org/0000-0002-9104-9680}{\textcolor[rgb]{0.00,1.00,0.00}{\sf ID}}}$
        \and
       \\Zhu-Fang Cui\thanksref{NJU,INP,eZFC}%
       $\,^{\href{https://orcid.org/0000-0003-3890-0242}{\textcolor[rgb]{0.00,1.00,0.00}{\sf ID}}}$
       \and
       Craig D.~Roberts\thanksref{NJU,INP,eCDR}%
       $\,^{\href{https://orcid.org/0000-0002-2937-1361}{\textcolor[rgb]{0.00,1.00,0.00}{\sf ID}}}$
       \and
       Chang Xu\thanksref{NJU,INP,eCX}%
       $\,^{\href{https://orcid.org/0000-0003-0683-4201}{\textcolor[rgb]{0.00,1.00,0.00}{\sf ID}}}$  
}

\thankstext{eHYX}{hyxing@smail.nju.edu.cn}
\thankstext{eZNX}{znxu@smail.nju.edu.cn}
\thankstext{eZFC}{phycui@nju.edu.cn}
\thankstext{eCDR}{cdroberts@nju.edu.cn}
\thankstext{eCX}{cxu@nju.edu.cn}

\authorrunning{Hui-Yu Xing \emph{et al}.} 

\institute{
School of Physics, Nanjing University, Nanjing, Jiangsu 210093, China \label{NJU}
           \and
\mbox{$\;$}Institute for Nonperturbative Physics, Nanjing University, Nanjing, Jiangsu 210093, China \label{INP}
            }

\date{2022 June 02}

\maketitle

\end{CJK}

\begin{abstract}
Using a symmetry-preserving regularisation of a vector$\times$vector contact interaction (SCI), we complete a systematic treatment of twelve semileptonic transitions with vector meson final states: $D\to \rho$, $D_{(s)}\to K^\ast$, $D_s\to \phi$, $B\to \rho$, $B_s\to K^\ast$, $B_{(s)}\to D_{(s)}^\ast$, $B_c \to B_{(s)}^\ast, J/\psi, D^\ast$; and thereby finalise a unified analysis of semileptonic decays of heavy+heavy and heavy+light pseudoscalar mesons to both pseudoscalar and vector meson final states.
The analysis is marked by algebraic simplicity, few parameters, and the ability to consistently describe systems from Nambu-Goldstone modes to heavy+heavy mesons.
Regarding the behaviour of the transition form factors, the SCI results compare well wherever sound experimental or independent theory analyses are available; hence, the SCI branching fraction predictions should be a reasonable guide.
Considering the ratios $R(D_{(s)}^{(\ast)})$, $R(J/\psi)$, $R(\eta_c)$, whose values are key tests of lepton universality in weak interactions, the SCI values agree with Standard Model predictions.
The $B_{(s)}\to D_{(s)}^\ast$ transitions are used to predict the precursor functions that evolve into the universal Isgur-Wise function in the heavy-quark limit, with results that conform with those from other sources where such are available.
The study also exposes effects on the transition form factors that flow from  interference between emergent hadron mass from the strong interaction and Higgs boson couplings via current-quark masses, including flavour symmetry violation.
\end{abstract}



\maketitle


\section{Introduction}
\label{Sec:Introduction}
%
%
Nature has two known mechanisms for mass generation.  In connection with quantum chromodynamics (QCD), that associated with the Higgs boson (HB) is responsible for the current-quark masses, which range from a few MeV for the lightest quarks to nearly 200\,GeV for the top quark.  Within the Standard Model of particle physics (SM), each current mass is generated by a distinct HB coupling; so, one has a parametric representation but not a satisfying explanation for this hierarchy of scales \cite{Fritzsch:1999ee}.
The other source of mass appears to be a dynamical feature of QCD; namely, emergent hadron mass (EHM), which is thought to be responsible for an array of phenomena that include \cite{Roberts:2020hiw, Roberts:2021xnz, Aguilar:2021uwa, Binosi:2022djx} the generation of nuclear-size masses for baryons constituted from light quarks and nearly-massless pseudoscalar Nambu-Goldstone bosons whose existence is crucial to the stability of known nuclei.

Pseudoscalar mesons are special because they are massless in the absence of HB couplings.  Consequently, weak-interaction mixing between quark flavours, pa\-ra\-me\-trised using the Cabibbo-Kobayashi-Maskawa \linebreak[4] (CKM) matrix, links both Nature's sources of mass via the study of semileptonic decays of heavy+heavy and heavy+light pseudo\-scalar mesons.  In such transitions, the mass of the initial state owes to HB couplings and constructive EHM plus HB interference, with the following HB:EHM+HB mass budgets \cite{Xu:2021iwv}: $B_c(87:13)$, $B_{(s)}(80:20)$, $D_{(s)}(70:30)$.  When the final states are also pseudoscalar mesons, these mass budgets can be dramatically reversed, \emph{e.g}.\ \cite{Roberts:2021nhw}: $K(20:80)$, $\pi(5:95)$.  In contrast, vector meson mass budgets are more like those of baryons, being nonzero even in the absence of HB couplings, with EHM alone delivering the bulk of their masses in lighter-quark cases, \emph{e.g}., $\rho(97:3)$ \cite{Roberts:2020udq}.
Thus, whilst semileptonic pseudoscalar-to-pseudoscalar decays have received most attention, there are also good reasons to consider pseudoscalar-to-vector transitions that go beyond their use as independent constraints on the CKM matrix elements.

The typically large disparity in masses between initial and final states is a significant challenge in the study of heavy pseudoscalar meson semileptonic transitions.  No framework with a traceable link to QCD can directly surmount this difficulty today.  Nevertheless, using a variety of methods, attempts are being made for pseudoscalar \cite{Chang:1992pt, DelDebbio:1997ite, Liu:1997hr, AbdEl-Hady:1999jux, Colangelo:1999zn, Melikhov:2001zv, Lu:2002ny, Ebert:2003cn, Ebert:2003wc, Cheng:2003sm, Ball:2004ye, Wu:2006rd, Khodjamirian:2006st, Ivanov:2006ni, Lu:2007sg, Ivanov:2007cw, Barik:2009zz, Faustov:2013ima, Faustov:2014bxa, Xiao:2014ana, Wang:2015vgv, Shi:2016gqt, Lu:2018cfc, Gubernari:2018wyi, Ivanov:2019nqd, Hu:2019bdf, Zhang:2020dla, Gonzalez-Solis:2021pyh, Na:2011mc, Bouchard:2014ypa, Flynn:2015mha, FermilabLattice:2015mwy, Colquhoun:2016osw, Lubicz:2017syv, Monahan:2017uby, FermilabLattice:2019ikx, McLean:2019qcx, Cooper:2020wnj, Chakraborty:2021qav, Yao:2021pdy, Yao:2021pyf, Nayak:2021djn, Gao:2021sav} and vector \cite{Chang:1992pt, DelDebbio:1997ite, Liu:1997hr, AbdEl-Hady:1999jux, Colangelo:1999zn, Lu:2002ny, Ebert:2003cn, Cheng:2003sm, Ball:2004rg, Wu:2006rd, Khodjamirian:2006st, Ivanov:2006ni, Lu:2007sg, Ivanov:2007cw, Barik:2009zz, Faustov:2013ima, Faustov:2014bxa, Donald:2013pea, Xiao:2014ana, Sekihara:2015iha, Shi:2016gqt, Gubernari:2018wyi, Ivanov:2019nqd, Hu:2019bdf, Chang:2019mmh, Zhang:2020dla, Yao:2021pyf, Nayak:2021djn} final states.  Given the challenges involved, unified treatments of pseudoscalar and vector final states are of particular value.

Using a symmetry-preserving formulation of a vector $\times$ vector contact interaction (SCI) \cite{Roberts:2010rn, Roberts:2011wy}, a study of ten pseudoscalar-to-pseudoscalar semileptonic transitions was described in Ref.\,\cite{Xu:2021iwv}:
$D\to \pi$, $D_{(s)}\to K$, $B\to \pi$, $B_s\to K$, $B_{(s)}\to D_{(s)}$, $B_c \to B_{(s)}, \eta_c$.
Combined, these transitions provide information on four elements of the CKM matrix: $|V_{cd}|$, $|V_{cs}|$, $|V_{ub}|$, $|V_{cb}|$.  Herein, we extend that analysis to a large array of analogous transitions with vector meson final states, all sensitive to the same set of CKM matrix elements, \emph{viz}.\ the following twelve decays:
$D\to \rho$, $D_{(s)}\to K^\ast$, $D_s\to \phi$, $B\to \rho$, $B_s\to K^\ast$, $B_{(s)}\to D_{(s)}^\ast$, $B_c \to B_{(s)}^\ast, J/\psi, D^\ast$.

As Ref.\,\cite{Xu:2021iwv} provided a benchmark for Ref.\,\cite{Yao:2021pdy}, which used continuum Schwinger function methods (CSMs) to deliver a coherent, parameter-free treatment of pseu\-do\-scalar-to-pseudoscalar transitions, so will the current study serve for the extension of Ref.\,\cite{Yao:2021pdy} to vector final states.
This is because contemporary implementations of the SCI preserve the essence of more sophisticated treatments of the continuum bound-state problem yet introduce an algebraic simplicity.  Widespread use has revealed that, when interpreted judiciously, SCI predictions provide a valuable quantitative guide, see \emph{e.g}.\ Refs.\,\cite{Wang:2013wk, Segovia:2014aza, Xu:2015kta, Bedolla:2015mpa, Bedolla:2016yxq, Serna:2017nlr, Raya:2017ggu, Zhang:2020ecj, Yin:2021uom, Raya:2021pyr, Lu:2021sgg, Gutierrez-Guerrero:2021rsx}.
Thus, SCI results provide a means by which one may check the validity of algorithms employed in calculations that rely (heavily) upon high performance computing.

%
Section~\ref{SecTFFs} presents some general material on the structure of currents and definitions of form factors in semileptonic pseudoscalar-to-vector meson transitions.
Section~\ref{SecRL}, augmented by an extensive appendix, explains our implementation of the SCI, including constraint of the ultraviolet cutoff and determination of values for the inter\-action-dependent current-quark masses.
Section~\ref{SecSemiLepResults} reports our analysis of $D^0 \to \rho^-$, $D_s^+ \to K^{0\ast}$, $D^+ \to \bar K^{\ast 0}$, $D_s \to \phi$ transitions.
Section~\ref{SecSemiLepResultsBsc} describes $B\to \rho$, $B_s\to K^\ast$, $B_{(s)}\to D_{(s)}^\ast$, $B_c \to B_{(s)}^\ast, J/\psi, D^\ast$, including a discussion of the Isgur-Wise function.
Section~\ref{ESensitivity} considers the evolution of semileptonic transition form factors as the Higgs mechanism of current-quark mass generation becomes a more significant component of the final-state meson's mass, \emph{viz}.\ it discusses the issues of environment sensitivity and flavour symmetry breaking in electroweak transitions.
Section~\ref{epilogue} provides a summary and perspective.

\section{Transition Form Factors}
\label{SecTFFs}
Semileptonic decays of pseudoscalar mesons (P) to vector meson (V) final states are described by the following transition matrix element:
{\allowdisplaybreaks
\begin{align}
M_{\mu;\lambda}^{P\to V}&(P,Q) =\langle V(p_{V};\lambda) | \bar q_V i(\gamma_\mu - \gamma_\mu \gamma_5) q_P |P(k)\rangle \nonumber \\
& \hspace*{-1em} =  2 m_{V} \tfrac{Q_\mu \epsilon^\lambda\cdot Q}{Q^2} A_0(t)
+  [m_{P}+m_{V}]T_{\mu\nu}^Q \epsilon_\nu^\lambda \, A_1(t)  \nonumber \\
&  + [P_\mu + Q_\mu \tfrac{m_{P}^2-m_{V}^2}{Q^2}]
\frac{\epsilon^\lambda\cdot Q \, A_2(t)}{m_{P}+m_{V}} \nonumber \\
& 
+ \varepsilon_{\mu\nu\rho\sigma}\epsilon_\nu^\lambda k_\rho p_{V \sigma}
\frac{2 V(t)}{m_{P}+m_{V}}\,,
\label{EqMEs}
\end{align}
where $Q^2T_{\mu\nu}^Q = Q^2\delta_{\mu\nu} - Q_\mu Q_\nu$,
$P =k+p_{V}$,
$Q=k-p_{V}$,
with $k^2 = -m_{P}^2$, $p_{V}^2=-m_{V}^2$;
$\epsilon_\nu^\lambda(p_V)$ is a polarisation four-vector, with $\sum_{\lambda=1}^3\epsilon_\nu^\lambda(p_V)\epsilon_\mu^\lambda(p_V) = T_{\mu\nu}^{p_V}$;
the squared-momentum-transfer is $t=-Q^2$;
and
$t_\pm = (m_{P} \pm m_V)^2$.
($t_-$ is the largest accessible value of $t$ in the identified physical decay process.)
}
The scalar functions in Eq.\,\eqref{EqMEs} are the semileptonic transition form factors, which express all effects of hadron structure on the transitions.  
Ensuring the absence of kinematic singularities in Eqs.\,\eqref{EqMEs}, symmetries guarantee
\begin{align}
%
A_0(0) & = \tfrac{m_{P} + m_{V}}{2m_{V}} A_1(0) - \tfrac{m_{P} - m_{V}}{2m_{V}} A_2(0)\,.
\label{EQsymmetries}
\end{align}

Once results for the transition form factors are available, one can calculate the associated decay branching fractions from the differential decay width for $P \to V {\ell}^+ \nu_{\ell}$:
\begin{align}
\label{dGdt}
\left.\frac{d\Gamma}{dt}\right|_{P\to V{\ell}\nu_{\ell}} &
= \frac{G_F^2 |V_{q_Pq_V}|^2}{192\pi^3 m_{P}^3}
\lambda(m_{P},m_V,t) (1-\tfrac{m_{\ell}^2}{t})^2\, {\mathpzc H}^2,
\end{align}
where:
$G_F = 1.166 \times 10^{-5}\,$GeV$^{-2}$;
$|V_{q_Pq_V}|$ is the relevant CKM matrix element,
$\lambda(m_{P},m_V,t)^2 = (t_+ - t)(t_- - t)$;
\begin{align}
{\mathpzc H}^2 & = (H_+^2 + H_-^2 + H_0^2)(1+\tfrac{m_{\mathpzc l}^2}{2t})
+ \tfrac{3 m_{\mathpzc l}^2}{2 t} H_t^2,
\label{dGdtA}
\end{align}
$m_{\ell}^2\leq t \leq t_-$, $m_{\ell}$ is the lepton mass;
and
{\allowdisplaybreaks
\begin{subequations}
\label{Hpsi}
\begin{align}
\tfrac{1}{\surd t}H_\pm & = (m_{P}+m_{V})A_1 (t) \mp \frac{\lambda(m_{P},m_{V},t)}{m_{P}+m_{V}} V(t) \,,\\
H_0 & = \frac{1}{2 m_{V}} \left[
(m_{P}^2-m_{V}^2 -t)(m_{P}+m_{V}) A_1(t) \right.   \nonumber \\
& \quad \left. - \frac{\lambda(m_{P},m_{V},t)^2}{m_{P}+m_{V}} A_2(t)\right],\\
H_t & =\lambda(m_{P},m_{V},t)\, A_0(t)\,. \label{Htpsi}
\end{align}
\end{subequations}
It is plain from Eq.\,\eqref{dGdtA} that the contribution from $A_0(t)$ to any cross-section is kinematically suppressed.
After integrating Eq.\,\eqref{dGdt} to obtain the required partial widths, one quotes the branching fractions, ${\mathpzc B}_{P\to V{\ell} \nu_{\ell}}$, with respect to the total width determined from the pseudoscalar meson's lifetime \cite{Zyla:2020zbs}.
}

\section{Matrix Elements}
\label{SecRL}
When employing CSMs, the leading-order approximation to a matrix element like that in Eq.\,\eqref{EqMEs} is provided by the rainbow-ladder (RL) truncation \cite{Qin:2020rad}, which we choose to illustrate using the $D^0\to\rho^-$ transition:
\begin{align}
M_{\mu;\lambda}^{D^0\to \rho^-}&(P,Q)  = 2 N_c {\rm tr}\int\frac{d^4 t}{(2\pi)^4}
\Gamma_{D}(k) S_c(t+k) \nonumber \\
& \times i {\mathpzc W}_\mu^{cd}(Q) S_d(t+p) \Gamma_\rho(-p;\lambda) S_u(t)\,,
\label{dMD}
\end{align}
where $N_c=3$ and the trace is over spinor indices.
Three distinct types of matrix-valued function feature in Eq.\,\eqref{dMD}: propagators for the dressed-quarks involved in the transition process, here $S_f(t)$, $f=u,d,c$; Bethe-Salpeter amplitudes for the initial- and final-state me\-sons, $\Gamma_{M}$, $M=D^0 , \rho^-$; and the dressed $cd$ weak transition vertex, ${\mathpzc W}_\mu^{cd}$.
The $P\to V$ weak transition vertex has two pieces, \emph{viz}.\ vector and axial-vector:
\begin{equation}
\label{VminusA}
{\mathpzc W}_\mu^{cd} = {\mathpzc V}_\mu^{cd} - {\mathpzc A}_\mu^{cd}.
\end{equation}
Physically, the vector part, ${\mathpzc V}_\mu^{cd}$, must exhibit poles at $Q^2 + m_{D^\ast,D_{0}^\ast}^2=0$ and ${\mathpzc A}_\mu^{cd}$ has poles at $Q^2 + m_{D,  D_{1}}^2=0$.  The presence of such poles is a prerequisite for any valid analysis of $P\to V$ semileptonic transitions; and as will become apparent, they are manifest in our treatment.

The general structure of Eq.\,\eqref{dMD} is the same for any quark+antiquark interaction treated in RL truncation.  However, in writing Eq.\,\eqref{dMD}, we have implicitly assumed a SCI for all integral equations relevant to the problem.  This is signalled by the simplicity of the arguments of the Bethe-Salpeter amplitudes and weak transition vertex, as elucidated in \ref{AppendixSCI}.

\begin{table*}[t]
\caption{\label{fp0val}
\emph{Upper panel}\,--\,{\sf A}.
SCI predictions for almost all electroweak transition form factors can reliably be interpolated using Eq.\,\eqref{FormFt} and the coefficients in columns~1-12.
For $A_1(t)$ in $B^0\to\rho^-$, $B_s \to K^{\ast -}$, an accurate interpolation requires Eq.\,\eqref{FormFtB} instead, with $c= 0.64, 0.77$, respectively.
Column~13 lists $r_2=A_2(0)/A_1(0)$ and column~14, $r_V = V(0)/A_1(0)$.
\emph{Lower panel}\,--\,{\sf B}.
SCI computed branching fractions for kinematically allowed transitions (columns~1\,-\,3) and associated ratios (columns~4\,-\,5), obtained using experimentally determined masses, compared with empirical results \cite{Zyla:2020zbs, Aaij:2017tyk}, where available.
(All numerical entries should be multiplied by $10^{-3}$.)
Ref.\,\cite{Zyla:2020zbs} lists
$|V_{cd}| = 0.221(4)$, $|V_{cs}|= 0.987(11)$
$|V_{ub}| = 0.00382(24)$, $|V_{cb}|= 0.0410(14)$; and the following lifetimes (in seconds):
$\tau_{D^0} = 4.10 \times 10^{-13}$,
$\tau_{D^+} = 10.4 \times 10^{-13}$,
$\tau_{D_s^\pm} = 5.04 \times 10^{-13}$,
$\tau_{B^0} = 1.519 \times 10^{-12}$,
$\tau_{B_s^0} = 1.515 \times 10^{-12}$,
$\tau_{B_c^\pm} = 0.51 \times 10^{-12}$.
}
\begin{center}
\begin{tabular}{l|ccc|ccc|ccc|ccc|cc}\hline
 {\sf A}. & \multicolumn{3}{|c|}{$A_0(t)$} & \multicolumn{3}{|c|}{$A_1(t)$} & \multicolumn{3}{|c|}{$A_2(t)$} & \multicolumn{3}{|c|}{$V(t)$} &  \multicolumn{2}{|c}{$t=0$ ratios}  \\
 & $f_0$ & $a$ & $b$ & $f_0$ & $a$ & $b$ & $f_0$ & $a$ & $b$ & $f_0$ & $a$ & $b$ & $r_2$ & $r_V$ \\\hline
%
$\!\,1\;  D \rho$ & $0.61\,$ & $1.29\,$ & $\!\!0.27\!$ & $0.52\,$ & $0.15\phantom{3}\,$ & $\!\!-0.14\phantom{1}\!$ & $0.36\,$ & $0.60\,$ & $-0.042\!$ & $0.83\,$ & $0.87\,$ & $\phantom{-}0.0009\!$& $0.69\,$ & $1.58\,$ \\
$\!\,2\;  D_s \bar K^\ast$ & $0.62\,$ & $1.40\,$ & $\!\!0.27\!$ & $0.56\,$ & $0.22\phantom{3}\,$ & $\!\!-0.20\phantom{1}\!$ & $0.40\,$ & $0.72\,$ & $-0.047\!$ & $0.94\,$ & $0.98\,$ & $-0.0011\!$& $0.72\,$ & $1.68\,$ \\
$\!\,3\;  D \bar K^\ast$ & $0.68\,$ & $1.13\,$ & $\!\!0.14\!$ & $0.61\,$ & $0.18\phantom{3}\,$ & $\!\!-0.13\phantom{1}\!$ & $0.41\,$ & $0.56\,$ & $-0.038\!$ & $0.91\,$ & $0.82\,$ & $-0.0036\!$& $0.68\,$ & $1.50\,$ \\
$\!\,4\;  D_s \phi$ & $0.66\,$ & $1.28\,$ & $\!\!0.19\!$ & $0.61\,$ & $0.23\phantom{3}\,$ & $\!\!-0.17\phantom{1}\!$ & $0.44\,$ & $0.69\,$ & $-0.049\!$ & $1.00\,$ & $0.92\,$ & $-0.0042\!$& $0.72\,$ & $1.64\,$ \\
$\!\,5\;  B \rho$ & $0.38\,$ & $1.49\,$ & $\!\!0.42\!$ & $0.34\,$ & $0.20\phantom{3}\,$ & $\!\!\phantom{-}0.056\!$ & $0.32\,$ & $0.98\,$ & $-0.034\!$ & $0.45\,$ & $1.19\,$ & $0.10\;\!$& $0.95\,$ & $1.34\,$ \\
$\!\,6\;  B_s K^\ast$ & $0.36\,$ & $1.52\,$ & $\!\!0.41\!$ & $0.33\,$ & $0.26\phantom{3}\,$ & $\!\!-0.018\!$ & $0.31\,$ & $1.05\,$ & $-0.052\!$ & $0.46\,$ & $1.27\,$ & $0.13\;\!$& $0.95\,$ & $1.39\,$ \\
$\!\,7\;  B_c D^\ast$ & $0.28\,$ & $1.95\,$ & $\!\!0.51\!$ & $0.28\,$ & $0.063\,$ & $\!\!-2.24\phantom{3}\!$ & $0.28\,$ & $1.26\,$ & $-0.52\phantom{3}\!$ & $0.49\,$ & $1.64\,$ & $\phantom{0}0.023\;\!$& $1.00\,$ & $1.75\,$ \\
$\!\,8\;  B D^\ast$ & $0.74\,$ & $1.16\,$ & $\!\!0.11\!$ & $0.68\,$ & $0.42\phantom{0}\,$ & $\!\!-0.32\phantom{3}\!$ & $0.61\,$ & $0.98\,$ & $\phantom{-}0.011\!$ & $0.83\,$ & $1.12\,$ & $\phantom{0}0.078\;\!$& $0.90\,$ & $1.22\,$ \\
$\!\,9\;  B_s D_s^\ast$ & $0.64\,$ & $1.22\,$ & $\;\!0.099\!$ & $0.60\,$ & $0.46\phantom{0}\,$ & $\!\!-0.40\phantom{3}\!$ & $0.54\,$ & $1.03\,$ & $-0.021\!$ & $0.75\,$ & $1.17\,$ & $\phantom{0}0.059\;\!$& $0.91\,$ & $1.26\,$ \\
$\!\!\!10\; B_c J/\psi$ & $0.58\,$ & $1.56\,$ & $\!\!\!-0.045\,\!$ & $0.56\,$ & $0.61\phantom{0}\,$ & $\!\!-1.01\phantom{3}\!$ & $0.52\,$ & $1.24\,$ & $-0.43\!$ & $0.88\,$ & $1.46\,$ & \rule{-2ex}{0ex}$-0.19\;\!$& $0.93\,$ & $1.56\,$ \\
$\!\!\!11\; B_c B^\ast$ & $0.41\,$ & \rule{-2ex}{0ex}$20.4\,$ & \rule{-2ex}{0ex}$\!\!\!48.6\,\!$ & $0.43\,$ & \rule{-2ex}{0ex}$10.5\phantom{0}\,$ & \rule{-2ex}{0ex}$\!\!-56.6\phantom{3}\!$ & $0.72\,$ & \rule{-2ex}{0ex}$15.1\,$ & \rule{-0.5ex}{0ex}$10.8\!$ & $3.01\,$ & \rule{-2ex}{0ex}$15.3\,$ & \rule{-4ex}{0ex}$-12.2\;\!$& $1.66\,$ & $6.93\,$ \\
$\!\!\!12\; B_c B_s^\ast$ & $0.45\,$ & \rule{-2ex}{0ex}$18.0\,$ & \rule{-2ex}{0ex}$\!\!\!26.5\,\!$ & $0.47\,$ & \rule{0.2ex}{0ex}$9.95\phantom{0}\,$ & \rule{-2ex}{0ex}$\!\!-59.5\phantom{3}\!$ & $0.72\,$ & \rule{-2ex}{0ex}$14.5\,$ & \rule{+2.2ex}{0ex}$0.034\!$ & $3.11\,$ & \rule{-2ex}{0ex}$14.5\,$ & \rule{-4ex}{0ex}$-16.6\;\!$& $1.54\,$ & $6.65\,$ \\
\hline
\end{tabular}\vspace*{1em}

\begin{tabular}{l|ccc|cc|ccc|cc}\hline
$\!\!${\sf B}. & \multicolumn{5}{|c|}{SCI} &  \multicolumn{5}{|c}{\cite[PDG]{Zyla:2020zbs} or other, if indicated} \\\hline
 & $e \nu_e$ & $\mu \nu_\mu$ & $\tau \nu_\tau$ & $R_{\mu/e}$ & $R_{\tau/\mu}$
 & $e \nu_e$ & $\mu \nu_\mu$ & $\tau \nu_\tau$ & $R_{\mu/e}$ & \mbox{\rule{-5ex}{0ex}}$R_{\tau/\mu}$ \\\hline
$\,1\;  {\mathpzc B}_{D^0\rho^-}$ & $\phantom{1}1.27\,$ & $\phantom{1}1.21\,$ & & $0.95\,$ & & $\phantom{1}1.50(12)\,$ & & & & \\
$\,2\; {\mathpzc B}_{D^+\rho^0}$ & $\phantom{1}1.64\,$ & $\phantom{1}1.56\,$ & & $0.95\,$ & & $\phantom{1}2.18^{(17)}_{(25)}\,$ & $2.4(4)\!$ & & $1.12(19)\,$ & \\
$\,3\; {\mathpzc B}_{D_s\bar K^\ast}$ & $\phantom{1}1.76\,$ & $\phantom{1}1.68\,$ & & $0.95\,$ & & $\phantom{1}2.15(28)\,$ & & & & \\
$\,4\; {\mathpzc B}_{D^0 K^{\ast -}}$ & $21.0\phantom{0}\,$ & $19.7\phantom{0}\,$ & & $0.94\,$ & & $21.5(1.6)\,$ & $18.9(2.4)\!$ & & $0.88(13)\,$ & \\
$\,5\; {\mathpzc B}_{D^+\bar K^{\ast 0}}$ & $54.5\phantom{0}\,$ & $51.3\phantom{0}\,$ & & $0.94\,$ & & $54(1)\phantom{111}\,$ & $52.7(1.5)\!$ & & $0.98(3)\phantom{0}\,$ & \\
$\,6\; {\mathpzc B}_{D_s \phi}$ & $24.5\phantom{0}\,$ & $23.0\phantom{0}\,$ & & $0.94\,$ & & $\phantom{2}23.9(1.6)\phantom{1}\,$ & $19(5)\phantom{222}\!$ & & $\phantom{0}0.79(57)\phantom{0}\,$ & \\
$\,7\; {\mathpzc B}_{B^0 \rho^-}$ & $\phantom{00}0.445\,$ & $\phantom{00}0.443\,$ & $\phantom{0}0.232\,$ & $1.00\,$ & $0.52\,$ & $\phantom{00}0.294(21)\,$ & $\phantom{00}0.294(21)\!$ & & $1.0(1)\phantom{11}$ & \\
$\,8\; {\mathpzc B}_{B_s^0 K^{\ast -}}$ & $\phantom{00}0.468\,$ & $\phantom{00}0.466\,$ & $\phantom{0}0.249\,$ & $1.00\,$ & $0.53\,$ & &  & &  & \\
$\,9\; {\mathpzc B}_{B_c D^{\ast}}$ & $\phantom{00}0.143\,$ & $\phantom{00}0.142\,$ & $\phantom{0}0.076\,$ & $1.00\,$ & $0.54\,$ & &  & &  & \\
$\!\!10\; {\mathpzc B}_{B D^{\ast}}$ & $62.9\phantom{0}\,$ & $62.6\phantom{0}\,$ & $15.3\phantom{00}\,$ & $1.00\,$ & $0.24\,$ & $50.6(1.2)\!$ & $50.6(1.2)\,$ & \rule{-1ex}{0ex}$15.7(9)\,$ & $1.00(3)\,$ & \mbox{\rule{-7ex}{0ex}}$0.31(2)\phantom{0}\,$ \\
$\!\!11\; {\mathpzc B}_{B_s D_s^{\ast}}$ & $48.7\phantom{0}\,$ & $48.5\phantom{0}\,$ & $11.8\phantom{00}\,$ & $1.00\,$ & $0.24\,$ & & $54(5)\phantom{222}\!$ & &  & \\
$\!\!12\; {\mathpzc B}_{B_c J/\psi}$ & $15.8\phantom{0}\,$ & $15.7\phantom{0}\,$ & $3.62\phantom{00}\,$ & $0.99\,$ & $0.23\,$ & &  & & & \mbox{\rule{-3ex}{0ex}}$0.71(25)\,$\cite{Aaij:2017tyk}\\
%
$\!\!13\; {\mathpzc B}_{B_c B^\ast}$ & $\;1.33\,$ & $\;1.27\,$ & & $0.95\,$  & & &  & & & \\
%
$\!\!14\; {\mathpzc B}_{B_c B_s^\ast}$ & \rule{-1.8ex}{0ex}$17.5\,$ & \rule{-1.8ex}{0ex}$16.4\,$ & & $0.94\,$  & & &  & & & \\
\hline
\end{tabular}
\end{center}
\end{table*}

\section{Weak $D_{(s)}$ Semileptonic Transitions}
\label{SecSemiLepResults}
%
In the isospin-symmetry limit, there are four essentially distinct such processes: $D^0 \to \rho^-$, $D_s^+ \to K^{0\ast}$, $D^+ \to \bar K^{\ast 0}$, $D_s \to \phi$.  The first two measure $c\to d$ and the last two, $c\to s$.  Each provides information on the environmental sensitivity of these transitions.

On the physical domain, all $D_{(s)}$ transition form factors are monotonically increasing functions of $t$ that can reliably be interpolated using
\begin{equation}
\label{FormFt}
F(t) = f_0/[1 - a t/m_P^2 + b (t/m_P)^2]\,,
\end{equation}
where $m_P$ is the calculated mass of the initial-state pseudoscalar meson.  Our predictions for each of the transitions considered in this subsection are described by Eq.\,\eqref{FormFt} and the appropriate interpolation coefficients in Table~\ref{fp0val}A.

\begin{figure}[t]
\vspace*{2ex}

\leftline{\hspace*{0.5em}{\large{\textsf{A}}}}
\vspace*{-4ex}
\includegraphics[width=0.42\textwidth]{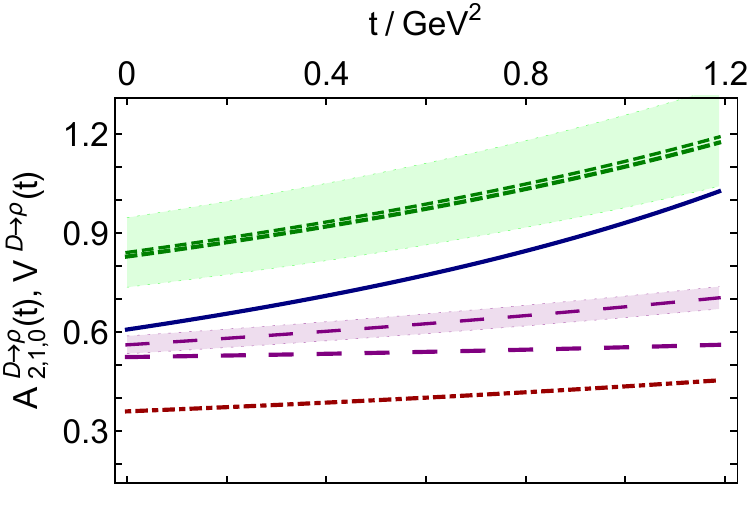}
\vspace*{-2ex}

\leftline{\hspace*{0.5em}{\large{\textsf{B}}}}
\vspace*{-4ex}
\includegraphics[width=0.42\textwidth]{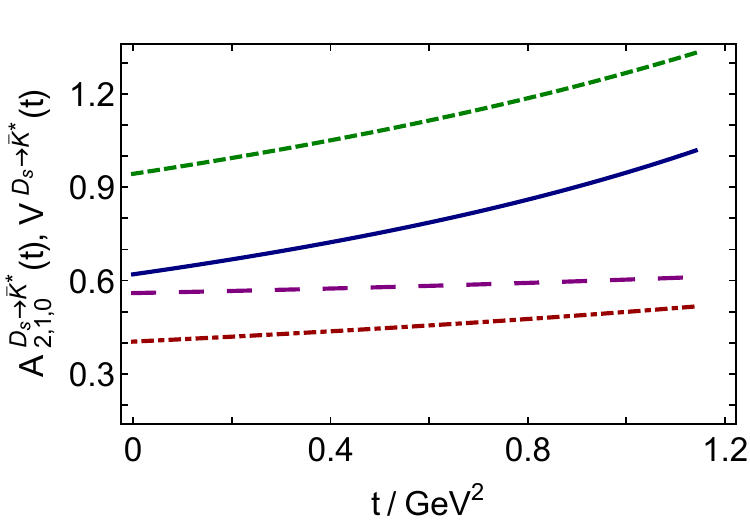}
\caption{\label{FigDrho}
{\sf Panel A}.
$D \to \rho$ transition form factors.  SCI predictions:
$A_2(t)$ -- dot-dashed red curve;
$A_1(t)$ -- long-dashed purple;
$A_0(t)$ -- solid blue;
$V(t)$ -- dashed green.
Thinner, like-texture and -colour curves within shaded bands are elementary monopole fits to data drawn from Ref.\,\cite[CLEO]{CLEO:2011ab}.
{\sf Panel B}.  SCI predictions for $D_s\to \bar K^\ast$ transition form factors, with associations as in Panel~A.  In this case, no comparable empirical form factors are available.
}
\end{figure}

\subsection{Cabibbo disfavoured $c\to d$ transitions}
Regarding $D^0 \to \rho^-$, data are available in Refs.\,\cite[CLEO]{CLEO:2011ab} and \cite[BESIII]{BESIII:2018qmf}.  Using contemporaneous values for $|V_{cd}|$ and $D$-meson lifetimes \cite{Nakamura:2010zzi} and fitting their data using elementary monopole functions, CLEO reported the following $t=0$ results
\begin{equation}
\begin{array}{ccc}
A_1 = 0.56(3)\,,\; & A_2=0.47(7)\,,\; & V=0.84(11)\,,\\
 & r_2 =0.83(12)\,,\; & r_V = 1.48(16)\,.\\
 \end{array}
\end{equation}
Following an analogous procedure, BESIII obtained consistent values for the $t=0$ ratios: $r_2 =0.845(69)$, $r_V = 1.695(97)$.
As evident in Table~\ref{fp0val}, our SCI predictions are compatible with these empirical results.

The SCI $D\to \rho$ transition form factors are depicted in Fig.\,\ref{FigDrho}A, wherein they are compared with monopole forms inferred from data \cite[CLEO]{CLEO:2011ab}.  Given the simplicity of those fits, the agreement is good.  Using the calculated form factors, Eq.\,\eqref{dGdt} yields the fractions in Table~\ref{fp0val}B\,-\,Rows~1-2.
Contrasted with model results in Ref.\,\cite{Ivanov:2019nqd}, the individual SCI branching fractions are $30$\% smaller.  On the other hand, those models produce $R_{\mu/e}=0.95(2)$, in agreement with the SCI prediction.  Combining SCI and model results \cite{Ivanov:2019nqd, Wu:2006rd, Sekihara:2015iha}, one obtains
\begin{equation}
\begin{array}{l|ccc}
10^{-3}                             & e\nu_e & \mu\nu_\mu & R_{\mu/e} \\\hline
{\mathpzc B}_{D^0\rho^-} &  1.66(30)  &   1.58(28) & 0.95(1)\\
{\mathpzc B}_{D^+\rho^0} &  2.14(38)  &  2.04(35) & 0.95(1) \\
\end{array}\,;
\end{equation}
outcomes which suggest that the Ref.\,\cite[PDG]{Zyla:2020zbs} value for ${\mathpzc B}_{D^+\rho^0\mu^+\nu_\mu}$ is too large.

Turning to $D_s^+ \to \bar K^\ast$, data on the $e^+\nu_e$ mode are available in Refs.\,\cite[CLEO]{CLEO:2009dyb, Hietala:2015jqa} and \cite[BESIII]{BESIII:2018xre}.  SCI predictions for the form factors are depicted in Fig.\,\ref{FigDrho}B.  They are reliably interpolated using Eq.\,\eqref{FormFt} with the coefficients in Table~\ref{fp0val}A\,-\,Row~2.  Little empirical information is available on the form factors; but Ref.\,\cite[BESIII]{BESIII:2018xre} reports $r_2=0.77(29)$, $r_V=1.67(38)$, values which agree with the SCI results in Table~\ref{fp0val}A\,-\,Row~2.

Our predictions for the branching fractions to the kinematically allowed semileptonic final states are listed in Table~\ref{fp0val}B\,-\,Row~2.  No experimental results are available for the $\mu^+\nu_\mu$ mode; but the SCI values are commensurate with model estimates collected in Ref.\,\cite{Ivanov:2019nqd}.  Combining those results \cite{Ivanov:2019nqd, Wu:2006rd, Sekihara:2015iha} with the SCI predictions, one obtains
\begin{equation}
\begin{array}{l|ccc}
 10^{-3}                                  & e\nu_e & \mu\nu_\mu & R_{\mu/e} \\\hline
{\mathpzc B}_{D_s\bar K^\ast} &  1.98(26)  &   1.88(26) & 0.95(1)\\
\end{array}\,.
\end{equation}

\subsection{Cabibbo favoured $c\to s$ transitions}
\begin{figure}[t]
\vspace*{2ex}

\leftline{\hspace*{0.5em}{\large{\textsf{A}}}}
\vspace*{-4ex}
\includegraphics[width=0.42\textwidth]{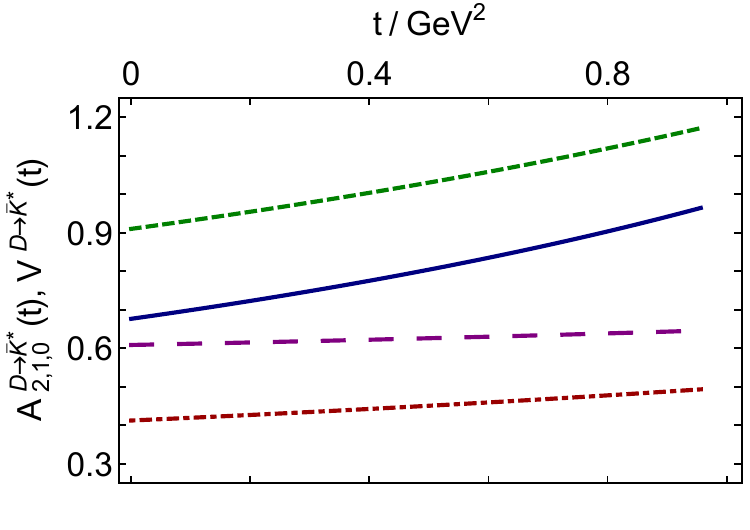}
\vspace*{-2ex}

\leftline{\hspace*{0.5em}{\large{\textsf{B}}}}
\vspace*{-4ex}
\includegraphics[width=0.42\textwidth]{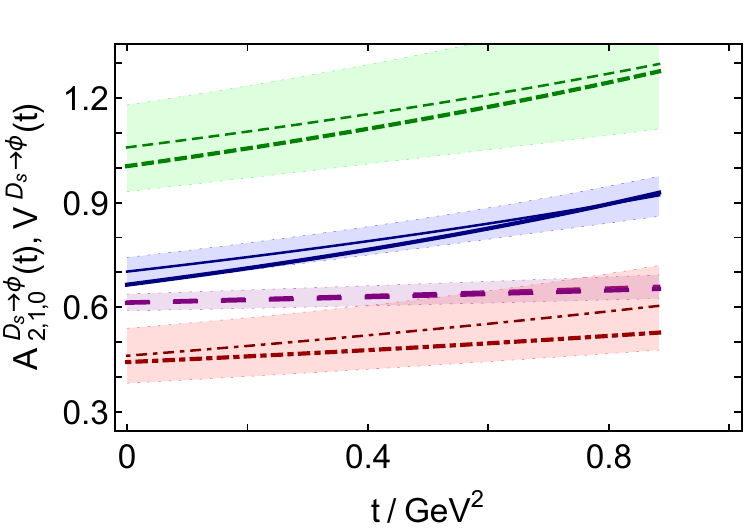}
\caption{\label{FigDKstar}
{\sf Panel A}.
SCI predictions for $D \to K^\ast$ transition form factors:
$A_2(t)$ -- dot-dashed red curve;
$A_1(t)$ -- long-dashed purple;
$A_0(t)$ -- solid blue;
$V(t)$ -- dashed green.
{\sf Panel B}.  SCI predictions for $D_s\to \phi$ transition form factors, with legend as in Panel~A.
Thinner, like-texture and -colour curves within shaded bands are lQCD results from Ref.\,\cite{Donald:2013pea}.
}
\end{figure}

SCI predictions for the $D^+ \to \bar K^\ast$ semileptonic transition form factors are drawn in Fig.\,\ref{FigDKstar}A.  They are reliably interpolated using Eq.\,\eqref{FormFt} with the coefficients in Table~\ref{fp0val}A\,-\,Row~3.  Regarding these form factors, some empirical information is available, \emph{e.g}., in Refs.\,\cite[FOCUS]{FOCUS:2004zbs} and \cite[BESIII]{BESIII:2018jjm}.   Using available data, Ref.\,\cite[PDG]{Zyla:2020zbs} compiles the following averages: $r_2=0.802(21)$, $r_V=1.49(05)$.  Whilst the value of $r_V$ agrees with the SCI result, that for $r_2$ is significantly larger.  In this connection, the SCI predictions are compatible with the model results collected in Ref.\,\cite{Ivanov:2019nqd}; and combined with those values, one obtains $r_2=0.64(24)$, $r_V=1.44(14)$.  Apparently, with $r_2$, there is some tension between experiment and theory.

Table~\ref{fp0val}B\,-\,Rows~4\,-\,5 list our predictions for the $D^{0} \to K^{\ast -}$, $D^{+} \to \bar K^{\ast 0}$ branching fractions to the allowed semileptonic final states.  They compare favourably with the averages reported in Ref.\,\cite[PDG]{Zyla:2020zbs}.  Combined with commensurate model estimates \cite{Wu:2006rd, Sekihara:2015iha}, one obtains
\begin{equation}
\begin{array}{l|ccc}
10^{-3}                                    & e\nu_e & \mu\nu_\mu & R_{\mu/e} \\\hline
{\mathpzc B}_{D^0 K^{\ast -}} &  21.2(0.3) &   19.9(2) & 0.94(1)\\
{\mathpzc B}_{D+ \bar K^{\ast 0}} &  54.6(1.0)  &   51.2(1) & 0.94(1)\\
\end{array}\,.
\end{equation}

Our $D_s \to \phi$ transition form factors are drawn in Fig.\,\ref{FigDKstar}B.  Reliable interpolations are obtained by using Eq.\,\eqref{FormFt} with the coefficients in Table~\ref{fp0val}A\,-\,Row~4.  Lattice QCD (lQCD) results for the form factors \cite{Donald:2013pea} are also drawn in Fig.\,\ref{FigDKstar}B.  They agree with our predictions.  This is further highlighted by a comparison between values computed at the maximum recoil point:
\begin{equation}
\begin{array}{l|ccc}
D_s \to \phi & r_2 & r_V & r_0 = A_0(0)/A_1(0) \\\hline
 \mbox{SCI} & \;0.72\phantom{(12)} & 1.64\phantom{(21)} & 1.08\phantom{(6)} \\
 \mbox{lQCD \cite{Donald:2013pea}} & \;0.74(12) & 1.72(21) & 1.14(6)
\end{array}\,.
\end{equation}
The results match well and also compare reasonably with the averages determined from available data \cite[PDG]{Zyla:2020zbs}: $r_2=0.84(11)$; $r_V=1.80(8)$.

Table~\ref{fp0val}B\,-\,Row~6 lists our predictions for the $D_s\to \phi$ branching fractions to the allowed semi\-leptonic final states.  They agree with the experimental results tabulated in Ref.\,\cite[PDG]{Zyla:2020zbs}.  However, there is only one measurement of the $\mu^+ \nu_\mu$ mode and it has a large error.  Given the agreement between SCI and lQCD determinations of the semileptonic form factors, there is merit in improving the precision of the $D_s \to \phi$ measurements.

\section{Weak $B_{(s,c)}$ Semileptonic Transitions}
\label{SecSemiLepResultsBsc}
%
In the isospin-symmetry limit, there are eight distinct processes, which we choose to be:
%
(\emph{i})
$B^0 \to \rho^-$,
$B_s^0 \to K^{\ast -}$;
$B_c^+ \to D^{\ast 0}$;
(\emph{ii}) $B^0 \to D^{\ast -}$,
$B_s^0 \to D_s^{\ast -}$,
$B_c^- \to J/\psi$;
(\emph{iii}) $B_c^+ \to B^{\ast 0}$;
and (\emph{iv}) $B_c^+ \to B_s^{\ast 0}$.
The first three measure $b\to u$ (\emph{i});
the next three, $b\to c$ (\emph{ii});
the sixth, $c\to d$ (\emph{iii}); and the last, $c\to s$ (\emph{iv}); plus, naturally, their environmental sensitivity.

All $B_{(s,c)}$ transition form factors are monotonically increasing functions of $t$ on their respective physical domains.  However, for decays involving a $\rho$ or $K^\ast$ in the final state, those domains can be large; consequently, we found that a reliable interpolation of $A_1(t)$ needs a $t^3$ term in the denominator:
\begin{equation}
\label{FormFtB}
F(t) = f_0/[1 - a t/m_P^2 + b (t/m_P)^2 - c(t/m_P)^3]\,,
\end{equation}
with $m_P$ the calculated mass of the initial state pseudoscalar meson.  Thus, whilst our predictions for almost all transition form factors considered in this subsection are well described by Eq.\,\eqref{FormFt} and the appropriate interpolation coefficients in Table~\ref{fp0val}A, there are two cases where Eq.\,\eqref{FormFtB} and $c\neq 0$ are necessary and we note this explicitly in the associated discussion.

\begin{figure}[t]
\vspace*{2ex}

\leftline{\hspace*{0.5em}{\large{\textsf{A}}}}
\vspace*{-4ex}
\includegraphics[width=0.42\textwidth]{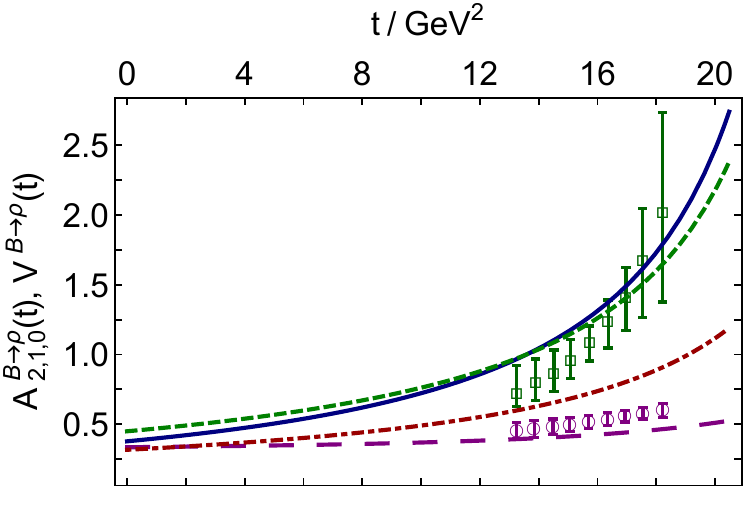}
\vspace*{-2ex}

\leftline{\hspace*{0.5em}{\large{\textsf{B}}}}
\vspace*{-4ex}
\includegraphics[width=0.42\textwidth]{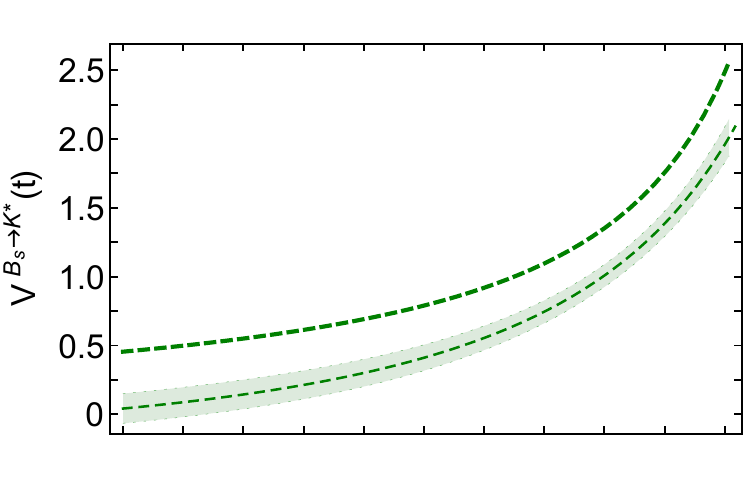}
\vspace*{-2ex}

\leftline{\hspace*{0.5em}{\large{\textsf{C}}}}
\vspace*{-4ex}
\includegraphics[width=0.42\textwidth]{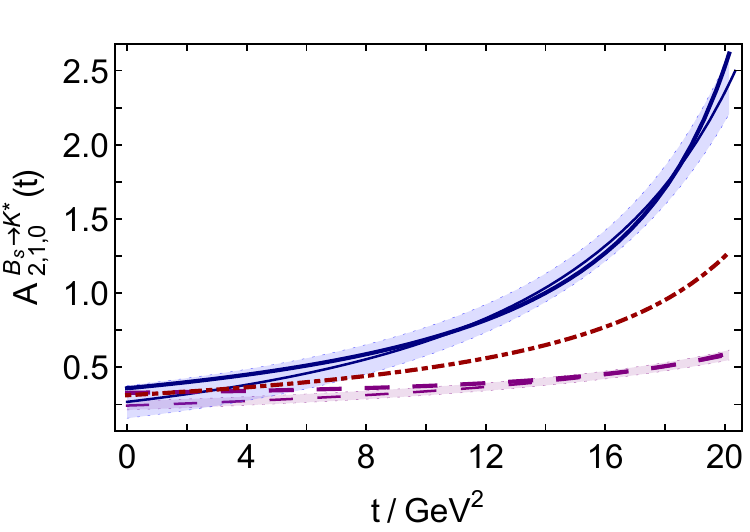}
\caption{\label{FigBrho}
{\sf Panel A}.
SCI predictions for $B \to \rho$ transition form factors:
$A_2(t)$ -- dot-dashed red curve;
$A_1(t)$ -- long-dashed purple;
$A_0(t)$ -- solid blue;
$V(t)$ -- dashed green.
Selected results from a simulation of quenched lQCD are also displayed \cite{Bowler:2004zb}:
$A_1(t)$ -- purple open circles;
$V(t)$ -- green open squares.
{\sf Panel B}.  SCI $B_s\to \bar K^\ast$ transition form factor $V(t)$  -- dashed green curve.
The thinner, like-textured and -coloured curve within the shaded band is a lQCD result from Ref.\,\cite{Horgan:2013hoa}.
{\sf Panel C}.  SCI $B_s\to \bar K^\ast$ transition form factors, $A_2(t)$, $A_1(t)$, $A_0(t)$ -- legend as in Panel A.
Thinner, like-texture and -colour curves within shaded bands are lQCD results for $A_1(t)$, $A_0(t)$ from Ref.\,\cite{Horgan:2013hoa}.
}
\end{figure}

\subsection{Cabibbo suppressed $b\to u$ transitions}
\label{SubSecbtou}
SCI predictions for $B^0 \to \rho^-$ semileptonic transition form factors are depicted in Fig.\,\ref{FigBrho}A.  Accurate interpolations are obtained using the coefficients in Table~\ref{fp0val}A\,-\,Row~5 ($c=0.64$ for $A_1$, otherwise $c=0$) in Eq.\,\eqref{FormFtB}.

Results from a simulation of quenched lQCD using unphysically large light-quark masses are available \cite{Bowler:2004zb}.  Those for $A_1(t)$, $V(t)$ are displayed in Fig.\,\ref{FigBrho}A in order to illustrate both their precision and $t$-domain coverage.  Nothing more recent is available from lQCD.

The SCI form factors in Fig.\,\ref{FigBrho}A yield the branching fractions to allowed semi\-leptonic final states listed in Table~\ref{fp0val}B\,-\,Row~7.  Where comparison is possible, the SCI results are roughly 50\% larger than those quoted in Ref.\,\cite[PDG]{Zyla:2020zbs}.
Notably, SCI predictions for $B^0 \to \pi^-$ exceed experiment by a factor of $\sim 2.8$ \cite[Table~3b]{Xu:2021iwv} because the maximum recoil value of the vector transition form factor is too large \cite[Table~3a]{Xu:2021iwv}.
We therefore compared the form factor maximum recoil values in Table~\ref{fp0val}A\,-\,Row~5 with the results in Refs.\,\cite{DelDebbio:1997ite, Lu:2002ny, Cheng:2003sm, Ball:2004rg, Lu:2007sg}, finding that the SCI values are 39(9)\% larger.  This is sufficient to explain the overestimated branching fractions.
Given the huge disparity in mass-scales between the initial and final states in the $B^0\to \rho^-$ transition, it is not too surprising that the SCI description is imperfect in this case.  Nevertheless, it is markedly better than the analogous treatment of $B^0\to \pi^-$; hence, the form factors should be a semiquantitatively sound guide, given that those for $B^0\to \pi^-$ are broadly compatible with experiment \cite[Fig.\,5]{Xu:2021iwv}.

The SCI $B_s^0 \to K^{\ast -}$ transition form factors are drawn in Figs.\,\ref{FigBrho}B, \ref{FigBrho}C.  Interpolations are obtained using \linebreak Eq.\,\eqref{FormFtB} and the coefficients in Table~\ref{fp0val}A\,-\,Row~6 ($c=0.77$ for $A_1$, otherwise $c=0$).

Results for $B_s\to K^\ast$ transition form factors are available from a simulation of unquenched lQCD, with four points on $t/{\rm GeV}^2 \in [14,19]$ at each of three pion masses, $m_\pi/{\rm GeV} = 0.31, 0.34, 0.52$, using lattice non-relativistic QCD to describe the $b$-quark \cite{Horgan:2013hoa}.
Those points were then fitted and extrapolated therein so as to arrive at estimates of the form factors on the entire kinematically accessible domain.  The fits for $V$, $A_1$, $A_0$ are drawn in Figs.\,\ref{FigBrho}B, \ref{FigBrho}C.  $A_2$ was not accessible in the lattice calculation.
Compared with the SCI prediction in Fig.\,\ref{FigBrho}B, the lQCD estimate for $V(t)$ is systematically lower on the entire $t$ domain.  The very low value of $V(0)=0.04(11)$ is a particular problem for the lQCD result.
There is better agreement between lQCD estimates and SCI predictions for $A_1(t)$, $A_0(t)$ -- Fig.\,\ref{FigBrho}C, although the lQCD curves again fall below on $t\lesssim 8\,$GeV$^2$.  Here it is worth recalling that all the lower-$t$ lQCD results involve a long reaching extrapolation from a small number of points on $t/{\rm GeV}^2 \in [14,19]$.

Working with the SCI form factors in Figs.\,\ref{FigBrho}B, \ref{FigBrho}C and the appropriate form of Eq.\,\eqref{dGdt}, one finds the branching fractions listed in Table~\ref{fp0val}B\,-\,Row~8.  No data are available on $B_s^0 \to K^{\ast -}$ semileptonic transitions.  In comparison with the results in Refs.\,\cite{Ball:2004rg, Lu:2007sg}, the SCI values for the form factors at the maximum recoil point are 35(21)\% larger.
Considering how a possible SCI overestimate of the $B \rho$ maximum recoil values might have affected the related branching fractions -- Table~\ref{fp0val}B\,-\,Row~7, one may estimate a correction to the results in Table~\ref{fp0val}B\,-\,Row~8 and therewith arrive at
\begin{equation}
\begin{array}{l|ccc}
          10^{-3}                         & e\nu_e & \mu\nu_\mu & \tau \nu_\tau \\\hline
{\mathpzc B}_{B_s K^{\ast -}}^{\rm rescaled} &  0.29 &   0.29 & 0.16\\
\end{array}\,.
\end{equation}

\begin{figure}[t]
\includegraphics[width=0.42\textwidth]{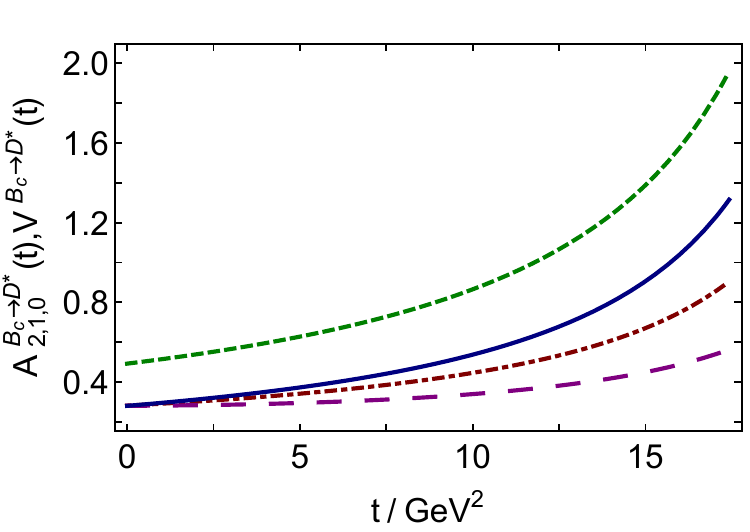}
\caption{\label{FigBcDast}
SCI predictions for $B_c^+ \to D^{\ast 0}$ transition form factors:
$A_2(t)$ -- dot-dashed red curve;
$A_1(t)$ -- long-dashed purple;
$A_0(t)$ -- solid blue;
$V(t)$ -- dashed green.
}
\end{figure}

SCI predictions for the $B_c^+ \to D^{\ast 0}$ semileptonic transition form factors are drawn in Fig.\,\ref{FigBcDast}.  In this case, interpolations are obtained using Eq.\,\eqref{FormFt} and the coefficients in Table~\ref{fp0val}A\,-\,Row~7.  The associated semileptonic branching fractions are listed in Table~\ref{fp0val}B\,-\,Row~9.
No data are available.
On the other hand, there are many calculations.  For instance, averaging the results in Refs.\,\cite{Chang:1992pt, Liu:1997hr, AbdEl-Hady:1999jux, Colangelo:1999zn, Ebert:2003cn, Nayak:2021djn} yields ${\mathpzc B}_{B_c \to D^{\ast}e^+\nu_e} = 0.14(5)$\permil, \emph{viz}.\ a central value that matches the SCI prediction.  Including the SCI result in the average, one finds
\begin{equation}
\bar{\mathpzc B}_{B_c \to D^{\ast}e^+\nu_e} = 0.14(4)\permil.
\end{equation}
Further, $R_{\tau/\mu}=0.58$ can be read from Ref.\,\cite{Ivanov:2006ni}, a value which complements the SCI result.
Notwithstanding these remarks, regarding $B_c^+ \to D^{\ast 0}$ semileptonic transitions, it is plain that theory precision can improve, calculations with closer links to QCD are necessary, and the want of data is sorely felt.

\subsection{Cabibbo inhibited $b\to c$ transitions: singly heavy}
\label{SubSecbtoc}
\begin{figure}[t]
\vspace*{2ex}

\leftline{\hspace*{0.5em}{\large{\textsf{A}}}}
\vspace*{-4ex}
\includegraphics[width=0.42\textwidth]{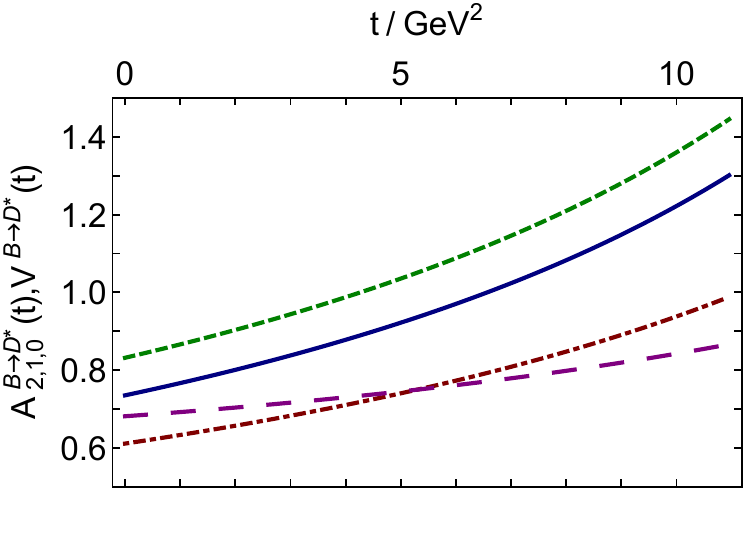}
\vspace*{-2ex}

\leftline{\hspace*{0.5em}{\large{\textsf{B}}}}
\vspace*{-4ex}
\includegraphics[width=0.42\textwidth]{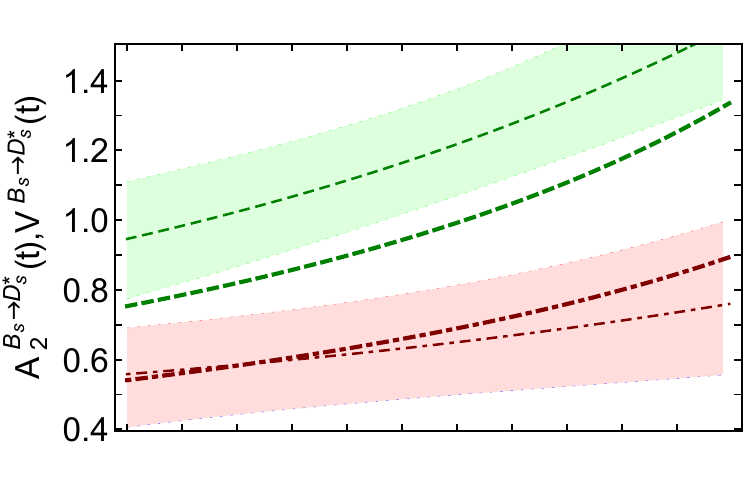}
\vspace*{-2ex}

\leftline{\hspace*{0.5em}{\large{\textsf{C}}}}
\vspace*{-4ex}
\includegraphics[width=0.42\textwidth]{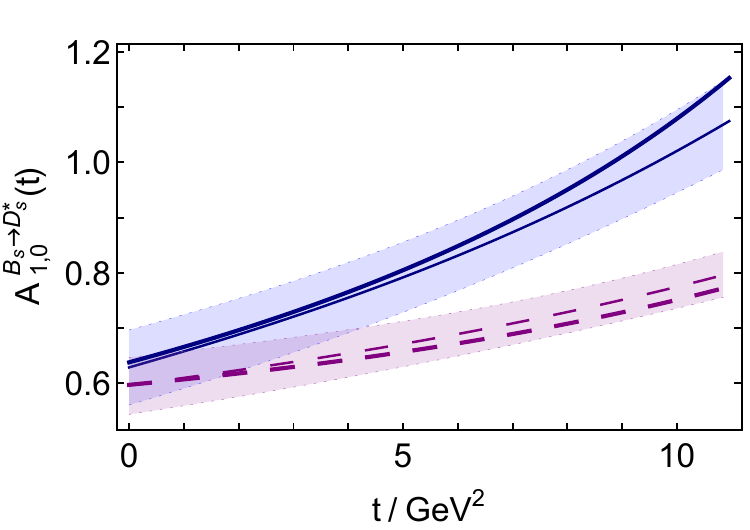}
\caption{\label{FigBDast}
{\sf Panel A}.
SCI predictions for $B^0 \to D^{\ast}$ transition form factors:
$A_2(t)$ -- dot-dashed red curve;
$A_1(t)$ -- long-dashed purple;
$A_0(t)$ -- solid blue;
$V(t)$ -- dashed green.
{\sf Panel B}.  $B_s\to D_s^{\ast}$ transition form factors:
$A_2(t)$ -- dot-dashed red curve;
$V(t)$ -- dashed green.
{\sf Panel C}.  $B_s\to D_s^{\ast}$ transition form factors:
$A_1(t)$ -- long-dashed purple curve;
$A_0(t)$ -- solid blue;
In Panels B and C, the thinner, like-texture and -colour curves within shaded bands are lQCD results from Ref.\,\cite{Harrison:2021tol}.
}
\end{figure}

In this sector, we first consider $B^0 \to D^{\ast -}$ semileptonic transitions.  The SCI results, drawn in Fig.\,\ref{FigBDast}A, are accurately interpolated using Eq.\,\eqref{FormFt} and the coefficients in Table~\ref{fp0val}A\,-\,Row~8.  They yield the branching fractions in Table~\ref{fp0val}A\,-\,Row~10.
Notably, whilst the $\tau \nu_\tau$ fraction is commensurate with the current data average \cite[PDG]{Zyla:2020zbs}, the $\ell\nu_\ell$, $\ell = e,\mu$, fractions are significantly larger.  Consequently, the value of $R_{\tau/\mu}$ -- often denoted $R(D^\ast)$ -- is of special interest.  The SCI result matches fairly with the value considered to be the SM prediction \cite{Fajfer:2012vx}: $0.252(3)$.  It therefore confirms a $2.9\sigma$ tension between theory and experiment on this ratio, which is a key test of lepton universality in Nature's weak interactions.

Analogous SCI results for $B_{(s)}^0 \to D_{(s)}$ transitions are \cite{Xu:2021iwv}:
\begin{equation}
R(D) = 0.27\,,\; R(D_s) = 0.26\,.
\end{equation}
The former compares well with the accepted SM prediction \cite{Fajfer:2012vx}: $R(D) = 0.297(17)$.  Empirically: \cite{HFLAV:2019otj}: $R(D)=0.34(3)$.  This value is 17\% smaller than the 2016 estimate $0.397 (49)$ \cite{HFLAV:2016hnz} and only $1.25\sigma$ larger than the SM result.

Similar tensions are discussed in connection with $B_c \to J/\psi$ semileptonic transitions \cite{Aaij:2017tyk, Yao:2021pyf}, to which we return in Sec.\,\ref{SubSecbtocHH}.

SCI predictions for the $B_s^0 \to D_s^{\ast -}$ semileptonic transition form factors are depicted in Figs.\,\ref{FigBDast}B, C.
They are reliably interpolated using Eq.\,\eqref{FormFt} and the coefficients in Table~\ref{fp0val}A\,-\,Row~9 and yield the branching fractions in Table~\ref{fp0val}A\,-\,Row~11.  Notably, each $B_s^0 \to D_s^{\ast -}$ fraction is uniformly 23\% smaller than the kindred $B^0 \to D^{\ast -}$ fraction.  This outcome is incompatible with the averages presented in Ref.\,\cite[PDG]{Zyla:2020zbs}.

On the other hand, there are many model studies of $B_{(s)}^0 \to D_{(s)}^{\ast -}$ semileptonic transitions and the results are widely scattered.  For instance, considering an analysis that uses two related methods in a simultaneous treatment of both sets \cite{Hu:2019bdf}, one finds
\begin{subequations}
\begin{align}
{\mathpzc B}_{B_{s}^0 \to D_{s}^{\ast -}\mu^+\nu_\mu}/{\mathpzc B}_{B^0 \to D^{\ast -}\mu^+ \nu_\mu} & = 1.0(3)\,,\\
{\mathpzc B}_{B_{s}^0 \to D_{s}^{\ast -}\tau^+\nu_\tau}/{\mathpzc B}_{B^0 \to D^{\ast -}\tau^+ \nu_\tau} & = 1.0(3)\,.
\end{align}
\end{subequations}
The SCI results fit well within these bands.  Moreover, profiting from the kindred study of $P\to P$ transitions \cite[Table~3]{Xu:2021iwv}, we arrive at the SCI prediction
\begin{equation}
{\mathpzc B}_{B_{s}^0 \to D_{s}^{-}\mu^+\nu_\mu}/
{\mathpzc B}_{B_{s}^0 \to D_{s}^{\ast -}\mu^+\nu_\mu} = 0.52\,,
\end{equation}
which is commensurate with a recent inference from measurements \cite[LHCb]{LHCb:2020cyw}: $0.46 \pm 0.013_{\rm stat} \pm 0.043_{\rm syst}$.

Lattice results for the $B_{(s)}^0 \to D_{(s)}^{\ast -}$ transition form factors are now available \cite{Harrison:2021tol}.  Curves developed from those results are drawn alongside the SCI predictions in Figs.\,\ref{FigBDast}B, \ref{FigBDast}C: within their uncertainties, the curves agree with the SCI predictions.   There is, perhaps, some tension between the results for the vector form factor: the $t=0$ value of the lQCD curve is unexpectedly large when compared with the many observable-constrained model calculations, \emph{e.g}., Refs.\,\cite{Faustov:2014bxa, Hu:2019bdf}.  Nevertheless, even in this case, the absolute relative difference between the SCI and lQCD curves is just 23(14)\%.

Such comparisons suggest that the ${\mathpzc B}_{B_{s}^0 \to D_{s}^{\ast -}\mu^+\nu_\mu}$ data average in Ref.\,\cite[PDG]{Zyla:2020zbs} deserves reconsideration.

\subsection{Isgur-Wise Function}
\label{Sec:IWF}
It is common to discuss $B_{(s)}^0 \to D_{(s)}^{\ast -}$ transitions in terms of form factors related to the Isgur-Wise function \cite{Isgur:1989ed}, defined as follows:
\begin{subequations}
\label{IWdefinition}
\begin{align}
h_{F}(w) & = F(t(w))/{\cal T}_+\,,\; F=A_0,A_2,V\,, \\
h_{A_1}(w) & = A_1(t(w)) 2 {\cal T}_+/[1+w]\,,
\end{align}
\end{subequations}
where  ${\cal T}_+=[m_P+m_V]/[2\sqrt{m_P m_V}]$;
$t(w)=m_P^2+m_V^2 - 2 m_P m_V w$;
and $1<w<[m_P^2+m_V^2]/[2m_P m_V]$, with $w=1$ corresponding to the zero recoil point.
This is because, in the true heavy-quark limit, $h_{A_0}(w) \equiv h_{A_1}(w)\equiv h_{A_2}(w)\equiv h_{V}(w)=:\xi(w)$, \emph{viz}.\ the transitions are described by a single, universal form factor with the property $\xi(1)=1$.
A similar statement is true for $B_{(s)}\to D_{(s)}$ transitions.

In the present cases, one cannot \emph{a priori} be certain that the conditions defining the heavy-quark symmetry limit are realised because the $c$-quark mass is not necessarily very much greater than the corresponding $u,d,s$-quark masses, \emph{e.g}., referring to \ref{AppendixSCI}\,-\,Table~\ref{Tab:DressedQuarks}, the $s/c$ ratio of dressed-quark masses is $\approx 1/3$.  Consequently, the degree to which the heavy-quark limit is accessible in Nature is an empirical question.

Regarding the transitions discussed in Sec.\,\ref{SubSecbtoc}, one finds the following zero recoil values of the functions in Eq.\,\eqref{IWdefinition}:
\begin{equation}
\begin{array}{l|cccc||lc}
 w=1 & h_{A_0} & h_{A_1} & h_{A_2} & h_{V} & w=1  & \xi_{f_+}\\\hline
B\to D^\ast & 0.86 & 0.94 & 1.12 & 1.25 & B\to D & 1.00 \\
B_s\to D_s^\ast & 0.79 & 0.85 & 1.02 & 1.18 & B_s\to D_s & 0.99 \\
\end{array}\,.
\end{equation}
In the last column, we have included results from the SCI analysis of $B_{(s)}\to D_{(s)}$ transitions \cite[Sec.\,7]{Xu:2021iwv}.

Results for $h_{A_1}(1)$ are available from lQCD \cite{McLean:2019sds}:
$B\to D^\ast = 0.914(24)$;
$B_s\to D_s^\ast = 0.902(13)$.
They are commensurate with the SCI values.
Evidently, the SCI predicts that with the $c$-quark involved in vector-meson final-states, there is a mean absolute relative deviation of 14(8)\% from the heavy-quark limit result $\xi(1)=1$.

\begin{figure}[t]
\vspace*{2ex}

\leftline{\hspace*{0.5em}{\large{\textsf{A}}}}
\vspace*{-4ex}
\includegraphics[width=0.42\textwidth]{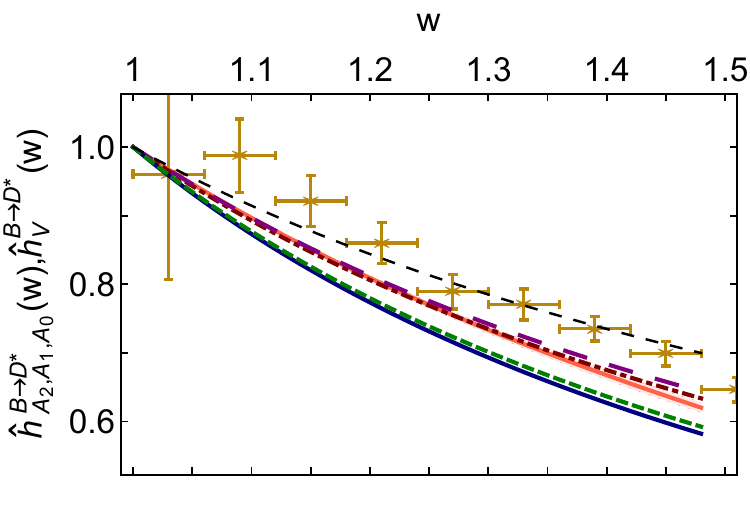}
\vspace*{-2ex}

\leftline{\hspace*{0.5em}{\large{\textsf{B}}}}
\vspace*{-4ex}
\includegraphics[width=0.42\textwidth]{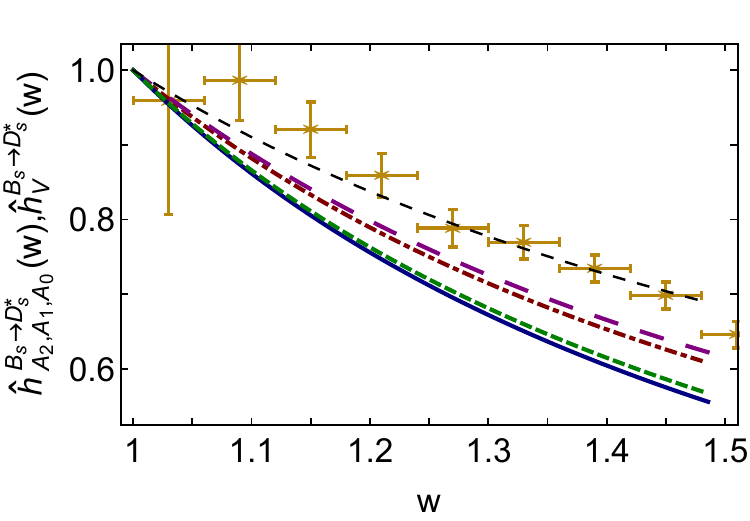}

\caption{\label{FigIWF}
{\sf Panel A}.
SCI predictions for precursors to the Isgur-Wise function, Eq.\,\eqref{IWdefinition}, obtained from $B^0 \to D^{\ast},D$ transitions:
$h_{A_2}(w)/h_{A_2}(1)$ -- dot-dashed red curve;
$h_{A_1}(w)/h_{A_1}(1)$ -- long-dashed purple;
$h_{A_0}(w)/h_{A_0}(1)$ -- solid blue;
$h_V(w)/h_V(0)$ -- dashed green;
$\xi_{f_+}(w)/\xi_{f_+}(1)$ -- mid-dashed, thin black curve \cite[Sec.\,7]{Xu:2021iwv}.
Empirical result \cite[Eqs.\,(177), (181)]{HFLAV:2019otj} -- thinner tomato-coloured curve within like-coloured bands.
{\sf Panel B}.  SCI predictions as in Panel A, but obtained using $B_s\to D_s^{\ast},D_s$ transitions.
Both panels: gold stars, Isgur-Wise function inferred from measurement of $B\to D$ \cite[Belle]{Glattauer:2015teq}.
}
\end{figure}

In Fig.\,\ref{FigIWF} we plot the precursor Isgur-Wise functions defined in Eq.\,\eqref{IWdefinition}, each divided by its zero recoil value.   All drawn curves can reliably be described using
\begin{equation}
\xi(w) = 1/[1-\rho^2 (w-1)]\,,
\end{equation}
with the following values of $\rho^2$:
\begin{equation}
\label{rho2values}
\begin{array}{l|cccc|c}
\rho^2 & \hat h_{V} & \hat h_{A_0} &\hat h_{A_1} & \hat h_{A_2} & \hat \xi_{f_+}\\\hline
B\to D^\ast & 1.42 & 1.48 & 1.16 & 1.20 & 0.91 \\
B_s\to D_s^\ast & 1.56 & 1.63 & 1.26 & 1.33 & 0.95 \\
\end{array}\,.
\end{equation}
Here, ``reliably'' means the mean absolute relative difference between a given function and the fit is $<0.2$\%.

The values of $\rho^2$ equate to the slope parameters of the precursor Isgur-Wise functions; and the results in Eq.\,\eqref{rho2values} emphasise that, in both sectors, $\hat h_{V}(w)\approx \hat h_{A_0}(w)$ and $\hat h_{A_1}(w)\approx \hat h_{A_2}(w)$.
The SCI values may be compared with a contemporary uncertainty-weighted average of $B\to D^\ast$ results \cite{HFLAV:2019otj}: $\rho^2_{A_1} = 1.122(24)$.  The associated empirical curve, drawn in Fig.\,\ref{FigIWF}A, agrees well with the SCI predictions for $\hat h_{A_1}(w)\approx \hat h_{A_2}(w)$.

\subsection{Cabibbo inhibited $b\to c$ transitions: doubly heavy}
\label{SubSecbtocHH}
\begin{figure}[t]
\vspace*{2ex}

\leftline{\hspace*{0.5em}{\large{\textsf{A}}}}
\vspace*{-4ex}
\includegraphics[width=0.42\textwidth]{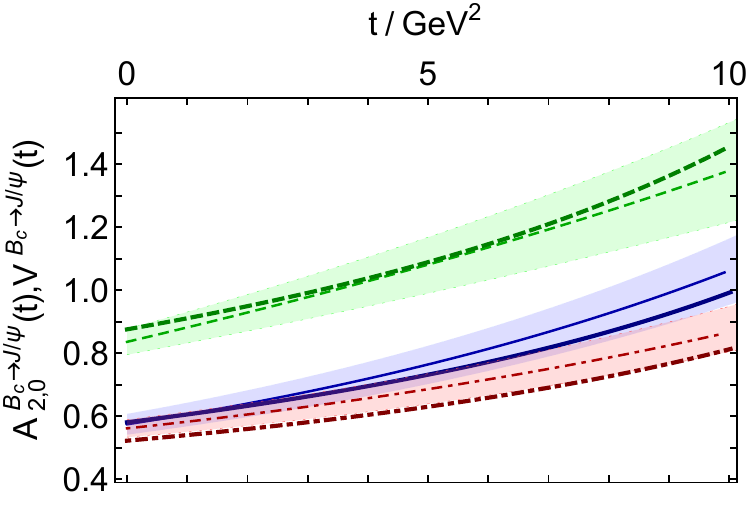}
\vspace*{-2ex}

\leftline{\hspace*{0.5em}{\large{\textsf{B}}}}
\vspace*{-4ex}
\includegraphics[width=0.42\textwidth]{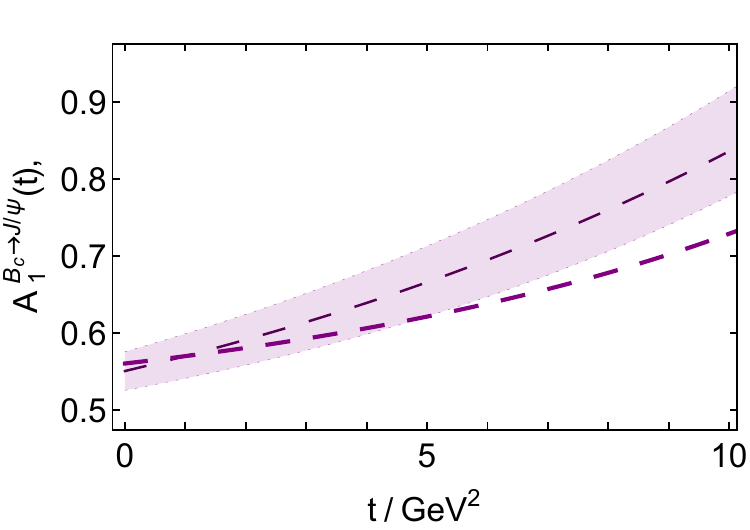}

\caption{\label{FigBcJ}
SCI predictions for $B_c^+\to J/\psi$ semileptonic transition form factors.
{\sf Panel A}.
$A_2(t)$ -- dot-dashed red curve;
$A_0(t)$ -- solid blue;
$V(t)$ -- dashed green.
{\sf Panel B}.
$A_1(t)$ -- long-dashed purple curve.
In both panels, the thinner, like-texture and -colour curves within shaded bands are the CSM predictions from Ref.\,\cite{Yao:2021pyf}.
}
\end{figure}

SCI results for the $B_c \to J/\psi$ semileptonic transition form factors are drawn in Fig.\,\ref{FigBcJ}, wherein they are compared with predictions obtained using a systematic, sym\-metry-preserving continuum approach to strong-inter\-ac\-tion bound-state problems \cite{Yao:2021pyf}.  Notwithstanding the SCI's algebraic simplicity, its results agree well with the sophisticated calculations: the mean error-weighted $\chi^2$ values are $0.14, 0.27, 1.25, 1.15$ for $V$, $A_0$, $A_2$, $A_1$, respectively.  Notably, too, the predictions in Ref.\,\cite{Yao:2021pyf} and the lQCD results in Ref.\,\cite{Harrison:2020gvo} agree within mutual uncertainties; hence, the SCI curves also agree with those obtained using lQCD.

The SCI form factors in Fig.\,\ref{FigBcJ} are accurately interpolated using Eq.\,\eqref{FormFt} with the coefficients in Table~\ref{fp0val}A\,-\,Row~10 and produce the branching fractions listed in Table~\ref{fp0val}B\,-\,Row~12.  Unsurprisingly, the SCI predictions agree with those in Ref.\,\cite{Yao:2021pyf}; hence, confirm the $2\sigma$ discrepancy between the SM prediction for $R(J/\psi)$ and the measurement in Ref.\,\cite[LHCb]{Aaij:2017tyk}.

The kindred SCI analysis of $B_c \to \eta_c$ transitions \cite{Xu:2021iwv} produces $R(\eta_c)=0.25$, a value 7\% larger than the SCI result for $R(J/\psi)$ in Table~\ref{fp0val}B\,-\,Row~12.  This size increase agrees semiquantitatively with that found in Ref.\,\cite{Yao:2021pyf}.

\subsection{Cabibbo disfavoured $c\to d$ transition}
\begin{figure}[t]
\vspace*{2ex}

\leftline{\hspace*{0.5em}{\large{\textsf{A}}}}
\vspace*{-4ex}
\includegraphics[width=0.42\textwidth]{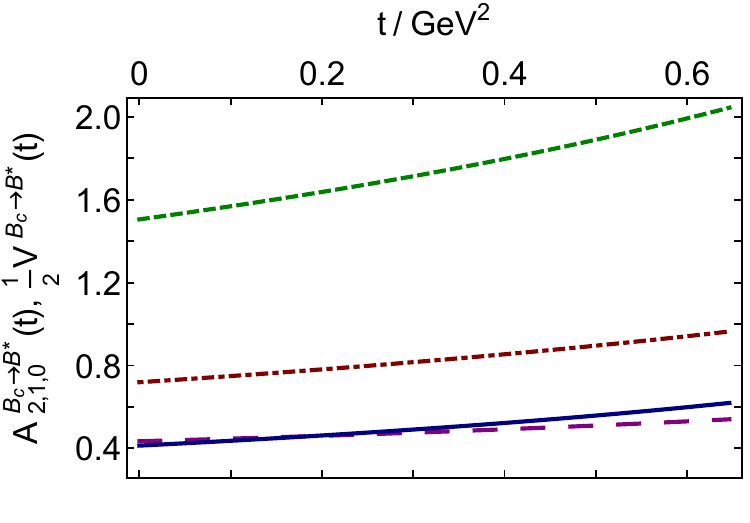}
\vspace*{-2ex}

\leftline{\hspace*{0.5em}{\large{\textsf{B}}}}
\vspace*{-4ex}
\includegraphics[width=0.42\textwidth]{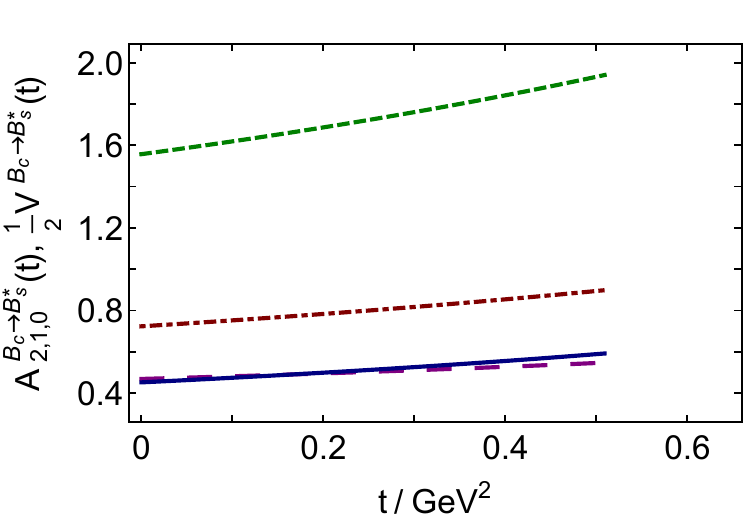}

\caption{\label{FigBcBast}
{\sf Panel A}.
SCI predictions for $B_c\to B^\ast$ transition form factors:
$A_2(t)$ -- dot-dashed red curve;
$A_1(t)$ -- long-dashed purple curve;
$A_0(t)$ -- solid blue curve;
$\tfrac{1}{2}V(t)$ -- dashed green curve.
{\sf Panel B}.  SCI predictions for$B_c\to B_s^\ast$ transition form factors, with legend as in Panel~A.
}
\end{figure}

SCI predictions for the $B_c^+ \to B^{\ast 0}$ transition form factors are depicted in Fig.\,\ref{FigBcBast}A.  They are interpolated using Eq.\,\eqref{FormFt} with the coefficients in Table~\ref{fp0val}A\,-\,Row~11.  In this instance, we have plotted $V(t)/2$ because the vector form factor is markedly enhanced compared with all axial form factors.  In fact, on the physically accessible domain, $V_{B_c B^\ast}(t)$ is significantly larger than all other form factors considered above.  This result is consistent with that found in Ref.\,\cite{Shi:2016gqt}.

The form factors in Fig.\,\ref{FigBcBast}A produce the branching fractions listed in Table~\ref{fp0val}B\,-\,Row~13.  No data are available.  On the other hand, the SCI value for the $e\nu_e$ final state is consistent with the results in Refs.\,\cite{Ebert:2003cn, Barik:2009zz, Shi:2016gqt}, the average of which is $1.32(12)$\permil.  Including the SCI value, the average becomes $1.32(9)$\permil.  Concerning the ratio of branching fractions, Ref.\,\cite{Shi:2016gqt} reports $0.95$, matching the SCI prediction.

\subsection{Cabibbo favoured $c\to s$ transition}
Our predictions for the $B_c^+ \to B_s^{\ast 0}$ transition form factors are drawn in Fig.\,\ref{FigBcBast}B.  The functions are accurately interpolated using Eq.\,\eqref{FormFt} with the coefficients in Table~\ref{fp0val}A\,-\,Row~12.  Here, for the same reasons as with $B_c^+ \to B^{\ast 0}$, we have drawn $V(t)/2$, whose magnitude and behaviour is similar to that found elsewhere \cite{Shi:2016gqt}.

Using the form factors in Fig.\,\ref{FigBcBast}B, one obtains the branching fractions listed in Table~\ref{fp0val}B\,-\,Row~14.  In the absence of data, we note that the SCI result for the $e\nu_e$ final state is consistent with the values in Refs.\,\cite{Ebert:2003cn, Barik:2009zz, Shi:2016gqt}, which average to $19.8(2.9)$\permil.  Including the SCI prediction in the average, one obtains $19.3(2.6)$\permil.  Regarding the ratio of branching fractions, Ref.\,\cite{Shi:2016gqt} reports $0.93$, matching the SCI prediction.

\section{Environment Sensitivity}
\label{ESensitivity}
%
%
%
Considering Nature's two mass generating mechanisms, it is worth considering the evolution of the form factors with increasing mass of the valence-quark that is a spectator to the transition; namely, as the current-mass contributed by the Higgs boson becomes a more significant part of the final-state meson's mass when compared with the EHM component.  Restricted to the $\{u, d, s\}$ quark sector, such effects express SU$(3)$ flavour symmetry violation.  Four classes can be defined for the semileptonic decays considered herein.
\begin{enumerate}[label=(\Roman*)]
\item $D \to\rho$, $D_s \to K^\ast$, $B_c\to B^\ast$:
    $c\to d$ transition with spectator quark, respectively, $u$, $s$, $b$.
    These are analogues of the pseudoscalar-to-pseudoscalar transitions
    $D\to\pi$, $D_s\to K$, $B_c\to B$.
\item $D\to K^\ast$, $D_s\to\phi$, $B_c \to B_s^\ast$:
    $c\to s$ transition with spectator quark, respectively, $u$, $s$, $b$; and analogues of $D\to K$, $B_c\to B_s$.
\item $B\to \rho$, $B_s\to K^\ast$, $B_c\to D^\ast$:
    $b\to u$ transition with spectator quark, respectively, $u$, $s$, $c$; and analogues of $B\to \pi$, $B_s \to K$.
\item $B\to D^\ast$, $B_s\to D_s^\ast$, $B_c\to J/\psi$:
    $b\to c$ transition with spectator quark, respectively, $u$, $s$, $c$;
    and analogues of $B\to D$, $B_s\to D_s$, $B_c\to \eta_c$.
\end{enumerate}

\begin{figure}[t]
\vspace*{2ex}

\leftline{\hspace*{0.5em}{\large{\textsf{A}}}}
\vspace*{-4ex}
\includegraphics[width=0.42\textwidth]{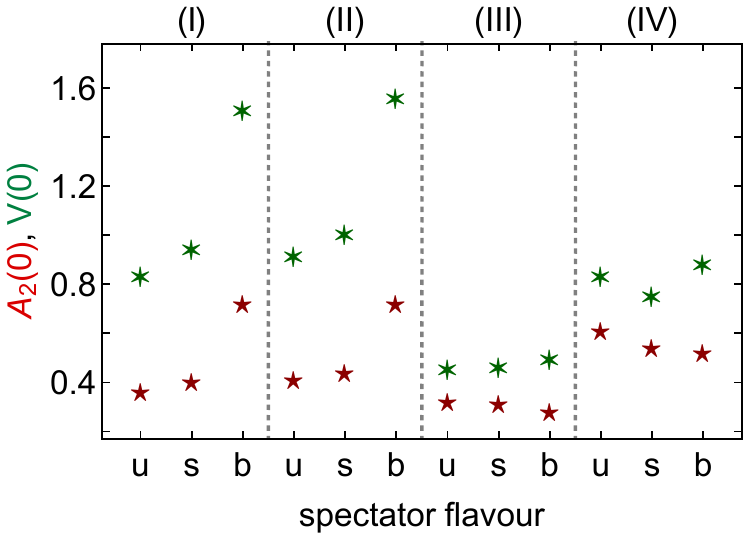}
\vspace*{1.5ex}

\leftline{\hspace*{0.5em}{\large{\textsf{B}}}}
\vspace*{-4ex}
\includegraphics[width=0.42\textwidth]{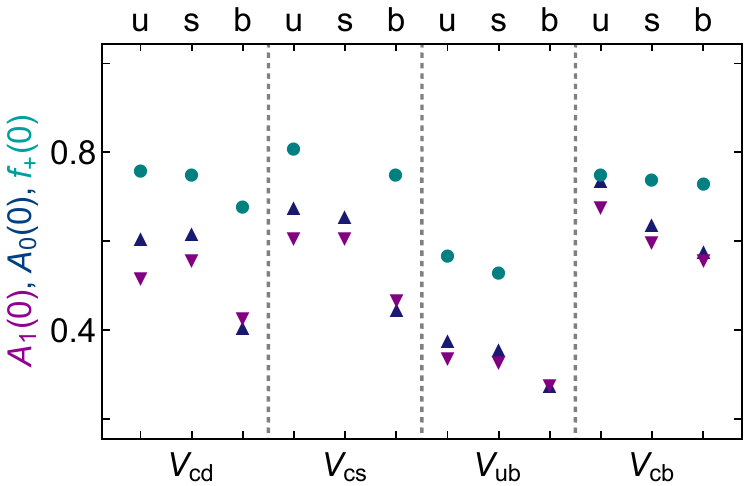}

\caption{\label{FigEnvironment}
SCI predictions for the transition form factor spectator current-mass dependence at maximum-recoil, labelled by the transition class defined when opening Sec.\,\ref{ESensitivity}.
{\sf Panel A}.
$A_2(0)$ -- red 5-point stars;
$V(0)$ -- green 6-points stars -- $\tfrac{1}{2}V(0)$ is plotted in the $b$-quark column for $V_{cd}$, $V_{cs}$.
{\sf Panel B}.
$A_1(0)$ -- purple down-triangles;
$A_0(0)$ -- blue up-triangles;
$f_+(0)$ in analogous pseudoscalar-to-pseudoscalar transitions \cite{Xu:2021iwv} -- teal circles.
}
\end{figure}

The SCI predictions split into two fairly distinct groups: Fig.\,\ref{FigEnvironment}A displays the results for $A_2(0)$, $V(0)$; and Fig.\,\ref{FigEnvironment}B, the values for $A_1(0)$, $A_0(0)$, and $f_+(0)$, the last being the vector form factor in analogous pseudoscalar-to-pseudoscalar transitions.
Regarding Fig.\,\ref{FigEnvironment}A, $V(0)$ \linebreak typically grows as the spectator mass increases.  $A_2(0)$ follows a similar pattern for $c\to d,s$ transitions -- Classes (I) and (II), but this is reversed for $b\to u,c$ -- (III) and (IV).
Turning to Fig.\,\ref{FigEnvironment}B, the general trend is for the zero-recoil value of each form factor to decrease with increasing spectator mass.
In the pseudoscalar-to-pseudoscalar case, the pattern in Fig.\,\ref{FigEnvironment}B can be understood by recalling that $f_+(t)$ is kindred to an elastic form factor when the current-quark masses in the initial- and final-states are similar.
No link to an elastic form factor can be made in any other case drawn in Fig.\,\ref{FigEnvironment}, so the behaviour expresses novel features of EHM-HB interference in electroweak transitions.

Restricting our considerations to cases with $u/d$, $s$ spectator quarks, the median absolute relative difference between maximum recoil values is $\underline{\rm ard} = 7.2$\% and the associated mean is $\overline{\rm ard} = 6.5(4.4)$\%.
Working with available information on leptonic decay constants -- Table~\ref{Tab:DressedQuarks}, the analogous results are $\underline{\rm ard} = 19$\% and $\overline{\rm ard} = 17(6)$\%, \emph{viz}.\ 2.7-times larger.

Notably, the Ward-Green-Takahashi identities satisfied by the two pieces of the weak transition vertex -- Ref.\,\cite[Eq.\,(29)]{Xu:2021iwv}  and Eq.\,\eqref{WGTI} -- involve differences and sums of quark current-masses multiplied by vertices whose structure can be greatly affected by EHM.  Consequently, one should anticipate complex interference effects.
The SCI predictions can be checked, \emph{e.g}., by expanding the array of transitions studied in Ref.\,\cite{Yao:2021pyf} and extending that analysis to vector-meson final-states.

\section{Summary and Perspective}
\label{epilogue}
Working with a symmetry-preserving regularisation of a vector$\times$vector contact interaction (SCI), we presented a unified treatment of twelve independent pseudoscalar-to-vector meson ($P\to V$) semileptonic transitions:
$D\to \rho$, $D_{(s)}\to K^\ast$, $D_s\to \phi$, $B\to \rho$, $B_s\to K^\ast$, $B_{(s)}\to D_{(s)}^\ast$, $B_c \to B_{(s)}^\ast, J/\psi, D^\ast$, each with four form factors;
and the masses and leptonic decay constants of twenty-three mesons that are either involved in the transitions or contribute in a material way to the weak transition vertex.
In completing this analysis, we simultaneously unified $P \to V$ semileptonic transitions with ten kindred pseudoscalar-to-pseudoscalar ($P \to P$) transitions, which were calculated elsewhere \cite{Xu:2021iwv} using precisely the same framework; and thereby finalised a comprehensive, coherent analysis of the semileptonic decays of heavy+heavy and heavy+light pseudoscalar mesons.

Our implementation of the SCI has four parameters, which are values of a mass-dependent quark+antiquark coupling strength chosen at the current-masses of the $u/d$, $s$, $c$, $b$ quarks.
The merits of the approach are its
algebraic simplicity;
paucity of parameters;
and simultaneous applicability to a wide variety of systems and processes, sometimes involving large disparities in mass-scales between initial and final states.

Concerning meson masses, which are long-wave\-length properties of the systems, the agreement between SCI predictions and experiment is good [Table~\ref{Tab:DressedQuarks}].  The comparison yields a median absolute relative difference $\underline{\rm ard} = 2$\% and mean $\overline{\rm ard} = 4$\%.  Including the leptonic decay constants, $\underline{\rm ard} = 3$\% and $\overline{\rm ard} = 10$\%.  Since leptonic decay constants are dominated by ultraviolet momenta, they are more of a challenge for the SCI.

Concerning the $t$ dependence of the $P\to V$ transition form factors, wherever experiment or solid theory results are available for comparison, the SCI results compare well [Secs.\,\ref{SecSemiLepResults}, \ref{SecSemiLepResultsBsc}].  The stiffness found with $P \to P$ transitions is largely avoided in $P \to V$ form factors because the SCI does not support a tensor term in the vector-meson Bethe-Salpeter amplitude when the quark+antiquark scattering kernel is treated in rainbow-ladder truncation.  The least satisfactory comparisons are found with $B\to \rho$ and $B_s\to K^\ast$; but owing to the huge disparity between mass-scales of the initial and final states, these transitions present difficulties for all available methods.  In every other case, our analysis suggests that the SCI branching fraction predictions should be a sound guide.  Of special interest are the SCI results for the branching fraction ratios $R(D_{(s)}^{(\ast)})$, $R(J/\psi)$, $R(\eta_c)$, whose values are key tests of lepton universality in Nature's weak interactions.  In all cases, the SCI values agree with accepted Standard Model predictions.

Working with $B_{(s)}\to D_{(s)}^\ast$ transitions, we provided predictions for each of the functions that evolves into the universal Isgur-Wise function in the heavy-quark limit [Sec.\,\ref{Sec:IWF}] and combined those results with the analogous functions in $B_{(s)}\to D_{(s)}$ transitions \linebreak[4]
[Fig.\,\ref{FigIWF}].  The SCI predictions for zero-recoil values are commensurate with lattice-QCD calculations and the functional form matches the curve inferred from experiment.  Recapitulating an observation from Ref.\,\cite{Xu:2021iwv}, when compared with $B\to D$ data from the Belle Collaboration, the SCI curve produces $\chi^2/$datum$=1.9$.

Since the SCI delivers a qualitatively and semiquantitatively reliable expression of both Nature's mass generating mechanisms, \emph{i.e}., emergent hadron mass and Higgs boson effects, it enables a comparison between the impacts of these effects on the entire array of electroweak transitions [Fig.\,\ref{FigEnvironment}].  Exploiting this and focusing on the evolution of the maximum-recoil value of each form factor as the mass of the spectator quark, $m_{\rm spec}$, is varied, we found that some patterns emerged.  For instance, $V(0)$ tends to increase as $m_{\rm spec}$ increases, whereas $A_{1,0}(0)$, $f_+(0)$ decrease; and $A_2(0)$ increases for $c\to d,s$ transitions, but decreases for $b\to u,c$.  These responses are expressions of the information encoded in the Ward-Green-Takahashi identities satisfied by the two pieces of each weak transition vertex.

Our analysis has highlighted the need for an extension of studies using symmetry-preserving formulations of realistic interactions in the analysis of semileptonic transitions involving heavy+heavy and heavy+light pseudoscalar mesons in the initial state.  There is a scarcity of such theory in this area, especially for transitions with a vector meson in the final state.  Work is therefore underway to adapt the framework exploited in Refs.\,\cite{Yao:2021pyf} for use in treating $B_c \to B_{(s)}$, $B_c \to D$, $B_{(s)} \to D_{(s)}$, and all the transitions considered herein, with the exception of the $J/\psi$ final state, which was studied in Ref.\,\cite{Yao:2021pdy}.

\begin{acknowledgements}
We are grateful for comments from P.~Cheng and Z.-Q.~Yao.
Work supported by:
National Natural Science Foundation of China (NSFC grant no. 12135007);
and
Natural Science Foundation of Jiangsu Province (grant no.\ BK20220323).
\end{acknowledgements}

\appendix

\section{Contact Interaction}
\label{AppendixSCI}
The keystone for the continuum meson bound-state problem is the quark+antiquark scattering kernel; and in RL truncation, it can be written ($k = p_1-p_1^\prime = p_2^\prime -p_2$):
\begin{subequations}
\label{KDinteraction}
\begin{align}
\mathscr{K}_{\alpha_1\alpha_1',\alpha_2\alpha_2'}  & = {\mathpzc G}_{\mu\nu}(k) [i\gamma_\mu]_{\alpha_1\alpha_1'} [i\gamma_\nu]_{\alpha_2\alpha_2'}\,,\\
 {\mathpzc G}_{\mu\nu}(k)  & = \tilde{\mathpzc G}(k^2) T^k_{\mu\nu}\,.
\end{align}
\end{subequations}

The defining quantity is $\tilde{\mathpzc G}$.  After roughly twenty years of study, it is now known that, owing to the emergence of a gluon mass-scale in QCD \cite{Boucaud:2011ug, Aguilar:2015bud, Cui:2019dwv, Roberts:2021nhw}, $\tilde{\mathpzc G}$ is nonzero and finite at infrared momenta; hence, can be written as follows:
\begin{align}
\tilde{\mathpzc G}(k^2) & \stackrel{k^2 \simeq 0}{=} \frac{4\pi \alpha_{\rm IR}}{m_G^2}\,.
\end{align}
In QCD \cite{Cui:2019dwv}: $m_G \approx 0.5\,$GeV, $\alpha_{\rm IR} \approx \pi$.
Following Ref.\,\cite{Yin:2021uom}, we retain this value of $m_G$ and, profiting from the fact that a SCI cannot support relative momentum between bound-state constituents, simplify the tensor structure in Eqs.\,\eqref{KDinteraction}:
\begin{align}
\label{KCI}
\mathscr{K}_{\alpha_1\alpha_1',\alpha_2\alpha_2'}^{\rm CI}  & = \frac{4\pi \alpha_{\rm IR}}{m_G^2}
 [i\gamma_\mu]_{\alpha_1\alpha_1'} [i\gamma_\mu]_{\alpha_2\alpha_2'}\,.
 \end{align}

A rudimentary form of confinement is implemented in the SCI by including an infrared regularisation scale, $\Lambda_{\rm ir}$, when defining all integral equations relevant to bound-state problems \cite{Ebert:1996vx}.  This artifice excises momenta below $\Lambda_{\rm ir}$, thereby eliminating quark+antiquark production thresholds \cite{Krein:1990sf}.  The usual choice is $\Lambda_{\rm ir} = 0.24\,$GeV\,$=1/[0.82\,{\rm fm}]$ \cite{Roberts:2011wy}, \emph{i.e}., which relates to a length scale that is roughly the same as the proton radii \cite{Cui:2022fyr}.

All integrals appearing in SCI bound-state equations require ultraviolet regularisation in a step that breaks the link between infrared and ultraviolet scales that is characteristic of QCD.  The associated ultraviolet mass-scales, $\Lambda_{\rm uv}$, thereby become physical parameters, which may be interpreted as upper bounds on the domains whereupon distributions within the associated systems are effectively momentum-independent.  For instance, the $\rho$-meson is larger in size than the $B^\ast$-meson; hence, one should expect $1/\Lambda_{\rm uv}^\rho > 1/\Lambda_{\rm uv}^{B^\ast}$.  As explained elsewhere \cite{Gutierrez-Guerrero:2019uwa, Yin:2021uom, Xu:2021iwv} and sketched below, this observation leads to a completion of the SCI through introduction of a scale-dependent coupling.

{\allowdisplaybreaks
For a quark of flavour $f$, the SCI gap equation is
\begin{align}
\label{GapEqn}
S_f^{-1}(p)  & = i\gamma\cdot p +m_f \nonumber \\
& \quad + \frac{16 \pi}{3} \frac{\alpha_{\rm IR}}{m_G^2}
\int \frac{d^4q}{(2\pi)^4} \gamma_\mu S_f(q) \gamma_\mu\,,
\end{align}
where $m_f$ is the associated quark current-mass.  Using a Poincar\'e-invariant regularisation, the solution is
\begin{equation}
\label{genS}
S_f^{-1}(p) = i \gamma\cdot p + M_f\,,
\end{equation}
with $M_f$, the dynamically generated dressed-quark mass, obtained by solving
\begin{equation}
M_f = m_f + M_f\frac{4\alpha_{\rm IR}}{3\pi m_G^2}\,\,{\cal C}_0^{\rm iu}(M_f^2)\,,
\label{gapactual}
\end{equation}
where  ($\tau_{\rm uv}^2=1/\Lambda_{\textrm{uv}}^{2}$, $\tau_{\rm ir}^2=1/\Lambda_{\textrm{ir}}^{2}$)
\begin{align}
\nonumber
{\cal C}_0^{\rm iu}(\sigma) &=
\int_0^\infty\! ds \, s \int_{\tau_{\rm uv}^2}^{\tau_{\rm ir}^2} d\tau\,{\rm e}^{-\tau (s+\sigma)}\\
& =
\sigma \big[\Gamma(-1,\sigma \tau_{\rm uv}^2) - \Gamma(-1,\sigma \tau_{\rm ir}^2)\big],
\label{eq:C0}
\end{align}
Here, the ``iu'' superscript stresses that the function depends on both the infrared and ultraviolet cutoffs and
$\Gamma(\alpha,y)$ is the incomplete gamma-function.
In general, functions of the following type arise in solving SCI bound-state equations:
\begin{align}
%
%
%
\overline{\cal C}^{\rm iu}_n(\sigma) & = \Gamma(n-1,\sigma \tau_{\textrm{uv}}^{2}) - \Gamma(n-1,\sigma \tau_{\textrm{ir}}^{2})\,,
\label{eq:Cn}
\end{align}
${\cal C}^{\rm iu}_n(\sigma)=\sigma \overline{\cal C}^{\rm iu}_n(\sigma)$, $n\in {\mathbb Z}^\geq$.}

Pseudoscalar mesons are generated as quark+anti\-quark bound-states, $f\bar g$.  They are described by a Bethe-Salpeter amplitude, whose SCI form is \cite{Xu:2021iwv}:
\begin{align}
\Gamma_{P}(Q) = \gamma_5 \left[ i E_{P}(Q) + \frac{1}{2 M_{f g}}\gamma\cdot Q F_{P}(Q)\right]\,,
\label{PSBSA}
\end{align}
$M_{f g}= M_f M_{g}/[M_f + M_{g}]$, $Q$ is the bound-state's total momentum, $Q^2 = -m_{P}^2$, $m_{P}$ is the meson's mass.  Additional details concerning these states as described by the SCI are provided elsewhere \cite[Sec.\,2]{Xu:2021iwv}.  It is nevertheless useful to specify the kernel:
\begin{subequations}
\label{fgKernel}
\begin{eqnarray}
\nonumber
{\cal K}_{EE}^{P} &=&
\int_0^1d\alpha \bigg\{
{\cal C}_0^{\rm iu}(\omega_{f g}( \alpha, Q^2))  \\
&&+ \bigg[ M_f M_{g}-\alpha \hat\alpha Q^2 - \omega_{f g}( \alpha, Q^2)\bigg]\nonumber \\
&&
\quad \times
\overline{\cal C}^{\rm iu}_1(\omega_{f g}(\alpha, Q^2))\bigg\},\\
\nonumber
{\cal K}_{EF}^{P} &=& \frac{Q^2}{2 M_{f g}} \int_0^1d\alpha\, \bigg[\hat \alpha M_f+\alpha M_{g}\bigg]\\
&& \quad \times \overline{\cal C}^{\rm iu}_1(\omega_{f g}(\alpha, Q^2)),\\
{\cal K}_{FE}^{P} &=& \frac{2 M_{f g}^2}{Q^2} {\cal K}_{EF}^{P} ,\\
\nonumber
{\cal K}_{FF}^{P} &=& - \frac{1}{2} \int_0^1d\alpha\, \bigg[ M_f M_{g}+\hat\alpha M_f^2+\alpha M_{g}^2\bigg]\\
&& \quad \times \overline{\cal C}^{\rm iu}_1(\omega_{f g}(\alpha, Q^2))\,,
\end{eqnarray}
\end{subequations}
where ($\hat \alpha = 1-\alpha$)
\begin{align}
\omega_{f g}(\alpha,Q^2) &= M_f^2 \hat \alpha + \alpha M_{g}^2 + \alpha \hat\alpha Q^2\,.
\label{eq:omega}
\end{align}

In arriving at the kernel in Eq.\,\eqref{fgKernel}, we followed Ref.\,\cite{Xu:2021iwv} in using the SCI identity
\begin{equation}
\label{SCIidentity}
0 = \int_0^1 d\alpha\,
\left[ {\cal C}_0^{\rm iu}(w_{fg}) + {\cal C}_1^{\rm iu}(w_{fg}) \right]\,,
\end{equation}
which is a consequence of requiring that there are no quadratic or logarithmic divergences in the treatment of integral equations; namely, that shifting integration variables is permitted \cite{GutierrezGuerrero:2010md}.  This condition is kindred to those implemented via di\-men\-sio\-nal-regularisation.

The SCI Bethe-Salpeter amplitude for a vector meson with polarisation $\lambda$ is \cite{Roberts:2011wy}:
\begin{equation}
\Gamma_V^\epsilon(Q;\lambda):=
\epsilon^\lambda\cdot\Gamma_V(Q;\lambda) = \epsilon^\lambda\cdot \gamma E_V(Q)\,.
\end{equation}
It is normalised canonically by rescaling such that
\begin{equation}
\label{normcan}
1=\left. \frac{d}{d Q^2}\Pi_{V}(Z,Q)\right|_{Z=Q},
\end{equation}
where, with the trace over spinor indices and $t_+ = t+Q$:
\begin{align}
\Pi_{V}(Z,Q) & = 2 N_c \tfrac{1}{3}\sum_\lambda {\rm tr}_{\rm D} \!\! \int\! \frac{d^4t}{(2\pi)^4} \nonumber \\
& \times  \Gamma_V^\epsilon(-Z;\lambda)
 S_f(t_+) \, \Gamma_V^\epsilon(Z;\lambda)\, S_g(t)\,.
 \label{normcan2}
\end{align}

In terms of the appropriate canonically normalised Bethe-Salpe\-ter amplitudes, pseudoscalar and vector meson leptonic decay constants are given by the following formulae:
\begin{align}
f_{P} &= \frac{N_c}{4\pi^2}\frac{1}{ M_{f g}}\,
\big[ E_{P} {\cal K}_{FE}^{P} + F_{P}{\cal K}_{FF}^{P} \big]_{Q^2=-m_{P}^2}\,,\label{ffg}\\
f_V & = - E_V \frac{3 N_c m_G^2}{8\pi m_V} K_V(Q^2=-m_V^2)\,,
\end{align}
where
\begin{align}
K_V(Q^2)&= -\frac{2\alpha_{\rm IR}}{3\pi m_G^2} \int_0^1 d\alpha \big[
M_f M_g - M_f^2 \hat\alpha \nonumber \\
& \qquad - M_g^2\alpha-2 \alpha\hat\alpha Q^2\big]
\overline {\cal C}^{\rm iu}_1(w_{f\bar g})\,.
\end{align}
With our normalisation, the empirical value of the pion's leptonic decay constant is $f_\pi =0.092\,$GeV \cite{Zyla:2020zbs}.

Improving upon the SCI introduced in Ref.\,\cite{Roberts:2011wy}, \linebreak Ref.\,\cite{Xu:2021iwv} kept all light-quark parameter values therein but determined the $s$-quark current mass, $m_s$, and $K$-meson ultraviolet cutoff, $\Lambda_{\rm uv}^K$, through a least-squares fit to measured values of $m_K$, $f_K$ whilst imposing the relation:
\begin{equation}
\alpha_{\rm IR}(\Lambda_{\rm uv}^{K}) [\Lambda_{\rm uv}^{K}]^2 \ln\frac{\Lambda_{\rm uv}^{K}}{\Lambda_{\rm ir}}
=
\alpha_{\rm IR}(\Lambda_{\rm uv}^{\pi}) [\Lambda_{\rm uv}^{\pi}]^2 \ln\frac{\Lambda_{\rm uv}^{\pi}}{\Lambda_{\rm ir}}\,.
\label{alphaLambda}
\end{equation}
This procedure eliminated one parameter by imposing the physical constraint that any increase in the momentum-space extent of a hadron wave function should be matched by a reduction in the effective coupling between the constituents.  One useful consequence is that critical over-binding is avoided. The procedure was repeated for the $c$-quark/$D$-meson and $\bar b$-quark/$B$-meson.  The complete set of results is collected in Table~\ref{Tab:DressedQuarks}.

\begin{table}[t]
\caption{\label{Tab:DressedQuarks}
Couplings, $\alpha_{\rm IR}/\pi$, ultraviolet cutoffs, $\Lambda_{\rm uv}$, and current-quark masses, $m_f$, $f=u/d,s,c,b$, that deliver a good description of flavoured pseudoscalar meson properties, along with the dressed-quark masses, $M$, and pseudoscalar meson masses, $m_{P}$, and leptonic decay constants, $f_{P}$, they produce; all obtained with $m_G=0.5\,$GeV, $\Lambda_{\rm ir} = 0.24\,$GeV.
Empirically, at a sensible level of precision \cite{Zyla:2020zbs}:
$m_\pi =0.14$, $f_\pi=0.092$;
$m_K=0.50$, $f_K=0.11$;
$m_{D} =1.87$, $f_{D}=0.15$;
$m_{B}=5.30$, $f_{B}=0.14$.
%
(We assume isospin symmetry and list dimensioned quantities in GeV.)}
\begin{center}
\begin{tabular*}
{\hsize}
{
l@{\extracolsep{0ptplus1fil}}|
c@{\extracolsep{0ptplus1fil}}|
c@{\extracolsep{0ptplus1fil}}
c@{\extracolsep{0ptplus1fil}}
c@{\extracolsep{0ptplus1fil}}
|c@{\extracolsep{0ptplus1fil}}
c@{\extracolsep{0ptplus1fil}}
c@{\extracolsep{0ptplus1fil}}}\hline
& quark & $\alpha_{\rm IR}/\pi\ $ & $\Lambda_{\rm uv}$ & $m$ &   $M$ &  $m_{P}$ & $f_{P}$ \\\hline
$\pi\ $  & $l=u/d\ $  & $0.36\phantom{2}$ & $0.91\ $ & $0.0068_{u/d}\ $ & 0.37$\ $ & 0.14 & 0.10  \\\hline
$K\ $ & $\bar s$  & $0.33\phantom{2}$ & $0.94\ $ & $0.16_s\phantom{7777}\ $ & 0.53$\ $ & 0.50 & 0.11 \\\hline
$D\ $ & $c$  & $0.12\phantom{2}$ & $1.36\ $ & $1.39_c\phantom{7777}\ $ & 1.57$\ $ & 1.87 & 0.15 \\\hline
$B\ $ & $\bar b$  & $0.052$ & $1.92\ $ & $4.81_b\phantom{7777}\ $ & 4.81$\ $ & 5.30 & 0.14
\\\hline
\end{tabular*}
\end{center}
\end{table}

The evolution of $\Lambda_{\rm uv}$ with $m_P$ reported in Table~\ref{Tab:DressedQuarks} is described by the following interpolation $(s=m_{P}^2)$:
\begin{equation}
\label{LambdaIRMass}
\Lambda_{\rm uv}(s) = 0.306 \ln [ 19.2 + (s/m_\pi^2-1)/2.70]\,.
\end{equation}
Using this formula, then the associated coupling for a meson $H$ is obtained using Eq.\,\eqref{alphaLambda} with $[\Lambda_{\rm uv}^{K}]^2 \to m_H^2=s$.  The properties of any meson for which a mass estimate is available can subsequently be calculated by solving the associated Bethe-Salpeter equation with the thus prescribed cutoff and coupling, using the dressed-quark propagators already determined.
This leads to the pseudoscalar and vector meson masses and decay constants listed in Table~\ref{Tab:MesonSpectrum}.  Calculated axial-vector meson masses are also reported in Table~\ref{Tab:MesonSpectrum}, but the leptonic decay constants are omitted because empirical comparisons are unavailable.  Kindred SCI predictions for these quantities are reported elsewhere \cite[Table~1]{Yin:2021uom}.

Considering the absolute value of the relative differences between the SCI results in Table~\ref{Tab:DressedQuarks} and listed comparison values, the median is 3.3\% and the mean is 7.6\% with a standard deviation of 10\%.
A more detailed discussion is presented in connection with Fig.\,1 in Ref.\,\cite{Xu:2021iwv}.  Evidently, as noted therein, the SCI is not a precision tool for electroweak physics; but when employed judiciously, it is qualitatively reliable and usually a good quantitative guide.

\begin{table}[t]
\caption{\label{Tab:MesonSpectrum}
Calculated masses (column 1), Bethe-Salpeter amplitudes (columns 3 and 4), and decay constants (column 5) for a representative selection of mesons.
Measured masses (column 2) from Ref.\,\cite{Zyla:2020zbs};
entries marked by ``$\ast$'' in this column from Ref.\,\cite{Mathur:2018epb}.
Leptonic decay constants (column 6): measured values, where known, from Ref.\,\cite{Zyla:2020zbs}; and others, marked with ``$\ast$'', drawn from lQCD studies \cite{Zyla:2020zbs, Aoki:2019cca, McNeile:2012qf, Colquhoun:2015oha, Lubicz:2017asp}.
%
%
(Dimensioned quantities in GeV.
Underlined entries from Table~\ref{Tab:DressedQuarks}.)
}
\begin{tabular*}
{\hsize}
{c@{\extracolsep{0ptplus1fil}}
l@{\extracolsep{0ptplus1fil}}||
l@{\extracolsep{0ptplus1fil}}
l@{\extracolsep{0ptplus1fil}}|
l@{\extracolsep{0ptplus1fil}}
l@{\extracolsep{0ptplus1fil}}|
l@{\extracolsep{0ptplus1fil}}
l@{\extracolsep{0ptplus1fil}}
}
\hline
$J^{P}$ &      Meson               &     $m^{\rm CI}$     &        $m^{\rm e/l}$         &         $E$      &      $F$         &       $f^{\rm CI}$         &       $f^{\rm e/l}$   \\
\hline
$0^{-}$ & $\pi(u\bar{d})$          &      \underline{0.14}       &          0.14            &        3.59     &     0.47$\ $        &         \underline{0.10}          &           0.092         \\
        & $K(u\bar{s})$            &      \underline{0.50}       &          0.50            &        3.70     &     0.55$\ $        &         \underline{0.11}          &           0.11         \\

        & $D(u\bar{c})$            &     \underline{1.87}       &  1.87    &        3.25     &     0.39$\ $        &         \underline{0.15}        &           0.15(1)         \\
        & $D_{s}(s\bar{c})$        &      1.96       &          1.97            &        3.45     &     0.54$\ $        &         0.16          &           0.18         \\
        & $\eta_{c}(c\bar{c})$     &      2.90       &          2.98            &        3.74     &     0.90$\ $        &         0.20 &           0.24(1)         \\  
        & $B(u\bar{b})$            &      \underline{5.30}    &          5.30            &        2.98     &     0.18$\ $        &         \underline{0.14}      &           0.13$^\ast$         \\
        & $B_{s}(s\bar{b})$        &      5.38       &          5.37            &        3.26     &     0.27$\ $        &         0.16          &           0.16$^\ast$         \\
        & $B_{c}(c\bar{b})$        &      6.16       &          6.28            &        4.25     &     0.79$\ $        &         0.21          &           0.30$^\ast$         \\
        & $\eta_{b}(b\bar{b})$     &      9.30       &          9.40            &        4.54     &     1.21$\ $        &         0.41 &           0.41(1)         \\  
\hline
$1^{-}$ & $\rho(u\bar{d})$         &      0.93       &          0.78            &        1.53     &              &         0.13          &           0.15         \\
        & $K^{\ast}(u\bar{s})$     &      1.10       &          0.89            &        1.31     &               &         0.15          &           0.16         \\
        & $\phi(s\bar{s})$         &      1.22       &          1.02            &        1.31     &              &         0.15          &           0.17         \\
        & $D^{\ast}(u\bar{c})$     &      2.09       &          2.01            &        1.25     &               &         0.15          &           0.16(1)$^\ast$         \\
        & $D_{s}^{\ast}(s\bar{c})$ &      2.18       &          2.11            &        1.30     &              &         0.15          &           0.19(1)$^\ast$         \\
        & $J/\psi(c\bar{c})$         &      3.07       &          3.10            &        1.54     &              &         0.18          &           0.29         \\
        & $B^{\ast}(u\bar{b})$     &      5.36       &          5.33            &        1.26     &               &         0.13          &           0.13$^\ast$         \\
        & $B_{s}^{\ast}(s\bar{b})$ &      5.45       &          5.42            &        1.34     &               &         0.14          &           0.15$^\ast$         \\
        & $B_{c}^{\ast}(c\bar{b})$ &      6.24       &          6.33$^\ast$     &        1.97     &              &         0.20          &           0.30(1)$^\ast$         \\
\hline
$1^{+}$         & $D_{1}(u\bar{c})$        &      2.27       &          2.42            &        0.77     &               & & \\ 
        & $D_{s1}(s\bar{c})$       &      2.39       &          2.46            &        0.71     &               & & \\ 
        & $B_{1}(u\bar{b})$        &      5.50       &          5.73            &        0.80     &              &   & \\ 
        & $B_{s1}(s\bar{b})$       &      5.62       &          5.83            &        0.72     &               &   & \\ 
        & $B_{c1}(c\bar{b})$&      6.54       &          6.74$^\ast$           &        0.48     &               &   & \\ 
\hline
\end{tabular*}
\end{table}

Turning now to the electroweak vertex in Eq.\,\eqref{VminusA}, a detailed discussion of the vector part is contained in Ref.\,\cite[Sec.\,3.2]{Xu:2021iwv}.  Herein, therefore, we describe our SCI treatment of the axial-vector part.  In paralleling that of the vector vertex, it effectively serves to recapitulate that analysis.  We continue to use the $cd$ transition as our exemplar.

The axial-vector component of the $cd$ weak transition vertex satisfies a Ward-Green-Takahashi identity:
\begin{align}
\label{WGTI}
Q_{\mu} {\mathpzc A}_\mu^{cd}(Q) & + i (m_c+m_d) {\mathpzc P}^{cd}(Q) \nonumber \\
& = S_c^{-1}(t+Q) i \gamma_5 + i \gamma_5 S_d^{-1}(t)\,,
\end{align}
where ${\mathpzc P}^{cd}(Q)$ is an analogous Dirac-pseudoscalar vertex.  As with all $n$-point functions, caution must be exercised when formulating the SCI solution procedure for ${\mathpzc A}_\mu^{cd}(Q)$, ${\mathpzc P}^{cd}(Q)$.

{\allowdisplaybreaks
The two vertices in Eq.\,\eqref{WGTI} satisfy inhomogeneous Bethe-Salpeter equations, \emph{viz}.\ in RL truncation:
\begin{subequations}
\begin{align}
{\mathpzc A}_\mu^{cd}& (Q) = \gamma_5\gamma_\mu  - \frac{16}{3} \frac{\pi \alpha_{\rm IR}}{m_G^2} \nonumber \\
& \times
 \int\frac{d^4 t}{(2\pi)^4} \! \gamma_\alpha S_c(t+Q) {\mathpzc A}_\mu^{cd}(Q) S_d(t)\gamma_\alpha\,,
 \label{vectorIBSE}  \\
\nonumber
{\mathpzc P}^{cd}&(Q)  = \gamma_5 - \frac{16}{3} \frac{\pi \alpha_{\rm IR}}{m_G^2} \\
& \times  \int\! \frac{d^4t}{(2\pi)^4}
\gamma_\alpha S_c(t+Q) {\mathpzc P}^{cd}(Q) S_d(t) \gamma_\alpha\,. \label{scalarIBSE}
\end{align}
\end{subequations}
So long as the regularisation scheme is symmetry preserving, then the solutions are
{\allowdisplaybreaks
\begin{subequations}
\begin{align}
{\mathpzc A}_\mu^{cd}& (Q) = \gamma_5\gamma_\mu^T F_{\mathpzc A}^{cd}(Q^2) \nonumber \\
& \qquad + \gamma_5 \gamma_\mu^L \tilde F_{\mathpzc A}^{cd}(Q^2) + i Q_\mu \gamma_5  E_{\mathpzc A}^{cd}(Q^2)\,,
\label{genvector}\\
i {\mathpzc P}^{cd}&(Q) = i \gamma_5 E^{cd}_{\mathpzc P}(Q^2) + \gamma_5 \gamma_\mu \frac{Q_\mu}{2 M_{cd}} F^{cd}_{\mathpzc P}(Q^2) \,,
\label{genscalar}
\end{align}
\end{subequations}
where $Q_\mu \gamma_\mu^T = 0$, $\gamma_\mu^T+\gamma_\mu^L = \gamma_\mu$.
}

Considering Eq.\,\eqref{scalarIBSE}, one readily obtains the following algebraic equation for the two terms in the pseudoscalar vertex:
\begin{subequations}
\begin{align}
\left[\begin{array}{c}
E^{cd}_{\mathpzc P}(Q^2) \\
F^{cd}_{\mathpzc P}(Q^2)
\end{array}\right]
& = \left[I - {\cal K}\right]^{-1}
\left[\begin{array}{c}
1 \\
0
\end{array}\right]\,,\\
I - {\cal K} & =
\left[\begin{array}{cc}
1 & 0 \\
0 & 1
\end{array}
\right]
-
\frac{4\alpha_{\rm IR}}{3\pi m_G^2}\left[\begin{array}{cc}
{\cal K}_{EE}^{P} & {\cal K}_{EF}^{P} \\
{\cal K}_{FE}^{P} & {\cal K}_{FF}^{P}
\end{array}
\right]\,,
\end{align}
\end{subequations}
where the kernel elements are given in Eq.\,\eqref{fgKernel} with $f\to d$, $g\to c$.  As promised, straightforward calculation reveals a pole in $\left[I - {\cal K}\right]^{-1}$ at the mass of the pseudoscalar $D$-meson; hence, so does the complete $c d$ electroweak vertex.

Working with Eqs.\,\eqref{vectorIBSE}, \eqref{genvector} and continuing via analogy with Ref.\,\cite[Sec.\,3.2]{Xu:2021iwv}, one finds:
\begin{align}
\label{PTsu}
F_{\mathpzc A}^{cd}(Q^2) &=  \frac{1}{1+K_A^{cd}(Q^2)}\,,\\
\nonumber
K_A^{cd}(Q^2)&= \frac{2\alpha_{\rm IR}}{3\pi m_G^2} \int_0^1 d\alpha \big[
M_d M_c + M_d^2 \hat\alpha \\
& \qquad + M_c^2\alpha + 2 \alpha\hat\alpha Q^2\big]
\overline {\cal C}^{\rm iu}_1(w_{dc})\,.
\end{align}
Calculation reveals that $F_{\mathpzc A}^{cd}(Q^2)$ exhibits a pole at the SCI mass of the axial-vector $D_1$-meson.  Furthermore,
\begin{subequations}
\begin{align}
\left[\begin{array}{c}
\tilde F^{cd}_{\mathpzc A}(Q^2)\\
E^{cd}_{\mathpzc A}(Q^2)
\end{array}\right]
& = \left[I - \tilde{\cal K}\right]^{-1}
\left[\begin{array}{c}
1 \\
0
\end{array}\right]\,,\\
I - \tilde{\cal K} = &
\left[\begin{array}{cc}
1 & 0 \\
0 & 1
\end{array}
\right]
 -
\frac{4\alpha_{\rm IR}}{3\pi m_G^2}\left[\begin{array}{cc}
{\cal K}_{FF}^{P} & \kappa {\cal K}_{FE}^{P} \\
\tfrac{1}{\kappa} {\cal K}_{EF}^{P} & {\cal K}_{EE}^{P}
\end{array}
\right]\,,
\end{align}
\end{subequations}
where $\kappa=Q^2/[2M_{cd}]$ and, again, the basic kernel elements are given in Eq.\,\eqref{fgKernel} with $f\to d$, $g\to c$.
A little algebra now reveals that $[I - \tilde{\cal K}]^{-1}$ also exhibits a pole at the mass of the pseudoscalar $D$-meson.

Return now to Eq.\,\eqref{WGTI}.  This Ward-Green-Taka\-ha\-shi identity entails
\begin{subequations}
\begin{align}
\tilde F_{\mathpzc A}^{cd}(Q^2) & = 1 - \frac{m_c+m_d}{2 M_{cd}}F^{cd}_{\mathpzc P}(Q^2)\,, \label{F5test}\\
Q^2 E^{cd}_{\mathpzc A}(Q^2) & = M_c + M_d - (m_c+m_d) E^{cd}_{\mathpzc P}(Q^2)\,.
\label{E5test}
\end{align}
\end{subequations}
These identities are guaranteed so long as the regularisation ensures there are no logarithmic or quadratic divergences.  As noted above, this is typically achieved by using Eq.\,\eqref{SCIidentity}.

It is worth remarking that when refocused on the axial-vector vertex relevant for neutron $\beta$-decay and working in the chiral limit, the analogue of Eq.\,\eqref{E5test} entails that the $du$ transition vertex exhibits a pion pole at $Q^2=0$ whose residue is intimately connected with the dressed-quark mass.  This has long been a known feature of QCD \cite{Maris:1997hd}

Testing Eq.\,\eqref{F5test}, we find it satisfied with an absolute relative discrepancy of 0.5(0.6)\%.  Turning to \linebreak Eq.\,\eqref{E5test}, the result is 2.3(1.7)\%.  The nonzero values arise because whilst Eq.\,\eqref{SCIidentity} is sufficient to guarantee the identities when the couplings and cutoffs are the same for both quarks connected to the vertex, it requires minor adjustments when these quantities are allowed to be flavour dependent,  Eqs.\,\eqref{alphaLambda}, \eqref{LambdaIRMass}.  Nonetheless, all mismatches are small and well within the standard SCI error; hence, have no discernible impact on our SCI predictions.


\begin{thebibliography}{105}
\providecommand{\natexlab}[1]{#1}
\providecommand{\url}[1]{\texttt{#1}}
\providecommand{\urlprefix}{URL }
\expandafter\ifx\csname urlstyle\endcsname\relax
  \providecommand{\doi}[1]{doi:\discretionary{}{}{}#1}\else
  \providecommand{\doi}[1]{doi:\discretionary{}{}{}\begingroup
  \urlstyle{rm}\url{#1}\endgroup}\fi
\providecommand{\bibinfo}[2]{#2}

\bibitem[{Fritzsch and Xing(2000)}]{Fritzsch:1999ee}
\bibinfo{author}{H.~Fritzsch}, \bibinfo{author}{Z.-Z. Xing},
  \bibinfo{title}{{Mass and flavor mixing schemes of quarks and leptons}},
  \bibinfo{journal}{Prog. Part. Nucl. Phys.} \bibinfo{volume}{45}
  (\bibinfo{year}{2000}) \bibinfo{pages}{1--81}.

\bibitem[{Roberts(2020)}]{Roberts:2020hiw}
\bibinfo{author}{C.~D. Roberts}, \bibinfo{title}{{Empirical Consequences of
  Emergent Mass}}, \bibinfo{journal}{Symmetry} \bibinfo{volume}{12}
  (\bibinfo{year}{2020}) \bibinfo{pages}{1468}.

\bibitem[{Roberts(2021)}]{Roberts:2021xnz}
\bibinfo{author}{C.~D. Roberts}, \bibinfo{title}{{On Mass and Matter}},
  \bibinfo{journal}{AAPPS Bulletin} \bibinfo{volume}{31} (\bibinfo{year}{2021})
  \bibinfo{pages}{6}.

\bibitem[{Aguilar et~al.(2022)Aguilar, Ferreira, and
  Papavassiliou}]{Aguilar:2021uwa}
\bibinfo{author}{A.~C. Aguilar}, \bibinfo{author}{M.~N. Ferreira},
  \bibinfo{author}{J.~Papavassiliou}, \bibinfo{title}{{Exploring smoking-gun
  signals of the Schwinger mechanism in QCD}}, \bibinfo{journal}{Phys. Rev. D}
  \bibinfo{volume}{105}~(\bibinfo{number}{1}) (\bibinfo{year}{2022})
  \bibinfo{pages}{014030}.

\bibitem[{Binosi(2022)}]{Binosi:2022djx}
\bibinfo{author}{D.~Binosi}, \bibinfo{title}{{Emergent Hadron Mass in Strong
  Dynamics}}, \bibinfo{journal}{Few Body Syst.}
  \bibinfo{volume}{63}~(\bibinfo{number}{2}) (\bibinfo{year}{2022})
  \bibinfo{pages}{42}.

\bibitem[{Xu et~al.(2021)Xu, Cui, Roberts, and Xu}]{Xu:2021iwv}
\bibinfo{author}{Z.-N. Xu}, \bibinfo{author}{Z.-F. Cui}, \bibinfo{author}{C.~D.
  Roberts}, \bibinfo{author}{C.~Xu}, \bibinfo{title}{{Heavy + light
  pseudoscalar meson semileptonic transitions}}, \bibinfo{journal}{Eur. Phys.
  J. C} \bibinfo{volume}{81}~(\bibinfo{number}{12}) (\bibinfo{year}{2021})
  \bibinfo{pages}{1105}.

\bibitem[{Roberts et~al.(2021)Roberts, Richards, Horn, and
  Chang}]{Roberts:2021nhw}
\bibinfo{author}{C.~D. Roberts}, \bibinfo{author}{D.~G. Richards},
  \bibinfo{author}{T.~Horn}, \bibinfo{author}{L.~Chang},
  \bibinfo{title}{{Insights into the emergence of mass from studies of pion and
  kaon structure}}, \bibinfo{journal}{Prog. Part. Nucl. Phys.}
  \bibinfo{volume}{120} (\bibinfo{year}{2021}) \bibinfo{pages}{103883}.

\bibitem[{Roberts and Schmidt(2020)}]{Roberts:2020udq}
\bibinfo{author}{C.~D. Roberts}, \bibinfo{author}{S.~M. Schmidt},
  \bibinfo{title}{{Reflections upon the Emergence of Hadronic Mass}},
  \bibinfo{journal}{Eur. Phys. J. ST}
  \bibinfo{volume}{229}~(\bibinfo{number}{22-23}) (\bibinfo{year}{2020})
  \bibinfo{pages}{3319--3340}.

\bibitem[{Chang and Chen(1994)}]{Chang:1992pt}
\bibinfo{author}{C.-H. Chang}, \bibinfo{author}{Y.-Q. Chen},
  \bibinfo{title}{{The Decays of B(c) meson}}, \bibinfo{journal}{Phys. Rev. D}
  \bibinfo{volume}{49} (\bibinfo{year}{1994}) \bibinfo{pages}{3399--3411}.

\bibitem[{Del~Debbio et~al.(1998)Del~Debbio, Flynn, Lellouch, and
  Nieves}]{DelDebbio:1997ite}
\bibinfo{author}{L.~Del~Debbio}, \bibinfo{author}{J.~M. Flynn},
  \bibinfo{author}{L.~Lellouch}, \bibinfo{author}{J.~Nieves},
  \bibinfo{title}{{Lattice constrained parametrizations of form-factors for
  semileptonic and rare radiative B decays}}, \bibinfo{journal}{Phys. Lett. B}
  \bibinfo{volume}{416} (\bibinfo{year}{1998}) \bibinfo{pages}{392--401}.

\bibitem[{Liu and Chao(1997)}]{Liu:1997hr}
\bibinfo{author}{J.-F. Liu}, \bibinfo{author}{K.-T. Chao},
  \bibinfo{title}{{$B_c$ meson weak decays and CP violation}},
  \bibinfo{journal}{Phys. Rev. D} \bibinfo{volume}{56} (\bibinfo{year}{1997})
  \bibinfo{pages}{4133--4145}.

\bibitem[{Abd El-Hady et~al.(2000)Abd El-Hady, Munoz, and
  Vary}]{AbdEl-Hady:1999jux}
\bibinfo{author}{A.~Abd El-Hady}, \bibinfo{author}{J.~H. Munoz},
  \bibinfo{author}{J.~P. Vary}, \bibinfo{title}{{Semileptonic and nonleptonic
  B(c) decays}}, \bibinfo{journal}{Phys. Rev. D} \bibinfo{volume}{62}
  (\bibinfo{year}{2000}) \bibinfo{pages}{014019}.

\bibitem[{Colangelo and De~Fazio(2000)}]{Colangelo:1999zn}
\bibinfo{author}{P.~Colangelo}, \bibinfo{author}{F.~De~Fazio},
  \bibinfo{title}{{Using heavy quark spin symmetry in semileptonic $B_c$
  decays}}, \bibinfo{journal}{Phys. Rev. D} \bibinfo{volume}{61}
  (\bibinfo{year}{2000}) \bibinfo{pages}{034012}.

\bibitem[{Melikhov(2002)}]{Melikhov:2001zv}
\bibinfo{author}{D.~Melikhov}, \bibinfo{title}{{Dispersion approach to quark
  binding effects in weak decays of heavy mesons}}, \bibinfo{journal}{Eur.
  Phys. J. direct C} \bibinfo{volume}{4} (\bibinfo{year}{2002})
  \bibinfo{pages}{1--154}.

\bibitem[{Lu and Yang(2003)}]{Lu:2002ny}
\bibinfo{author}{C.-D. Lu}, \bibinfo{author}{M.-Z. Yang}, \bibinfo{title}{{B to
  light meson transition form-factors calculated in perturbative QCD
  approach}}, \bibinfo{journal}{Eur. Phys. J. C} \bibinfo{volume}{28}
  (\bibinfo{year}{2003}) \bibinfo{pages}{515--523}.

\bibitem[{Ebert et~al.(2003{\natexlab{a}})Ebert, Faustov, and
  Galkin}]{Ebert:2003cn}
\bibinfo{author}{D.~Ebert}, \bibinfo{author}{R.~N. Faustov},
  \bibinfo{author}{V.~O. Galkin}, \bibinfo{title}{{Weak decays of the $B_c$
  meson to charmonium and $D$ mesons in the relativistic quark model}},
  \bibinfo{journal}{Phys. Rev. D} \bibinfo{volume}{68}
  (\bibinfo{year}{2003}{\natexlab{a}}) \bibinfo{pages}{094020}.

\bibitem[{Ebert et~al.(2003{\natexlab{b}})Ebert, Faustov, and
  Galkin}]{Ebert:2003wc}
\bibinfo{author}{D.~Ebert}, \bibinfo{author}{R.~N. Faustov},
  \bibinfo{author}{V.~O. Galkin}, \bibinfo{title}{{Weak decays of the $B_c$
  meson to $B_s$ and $B$ mesons in the relativistic quark model}},
  \bibinfo{journal}{Eur. Phys. J. C} \bibinfo{volume}{32}
  (\bibinfo{year}{2003}{\natexlab{b}}) \bibinfo{pages}{29--43}.

\bibitem[{Cheng et~al.(2004)Cheng, Chua, and Hwang}]{Cheng:2003sm}
\bibinfo{author}{H.-Y. Cheng}, \bibinfo{author}{C.-K. Chua},
  \bibinfo{author}{C.-W. Hwang}, \bibinfo{title}{{Covariant light front
  approach for s wave and p wave mesons: Its application to decay constants and
  form-factors}}, \bibinfo{journal}{Phys. Rev. D} \bibinfo{volume}{69}
  (\bibinfo{year}{2004}) \bibinfo{pages}{074025}.

\bibitem[{Ball and Zwicky(2005{\natexlab{a}})}]{Ball:2004ye}
\bibinfo{author}{P.~Ball}, \bibinfo{author}{R.~Zwicky}, \bibinfo{title}{{New
  results on $B \to \pi, K, \eta$ decay form factors from light-cone sum
  rules}}, \bibinfo{journal}{Phys. Rev. D} \bibinfo{volume}{71}
  (\bibinfo{year}{2005}{\natexlab{a}}) \bibinfo{pages}{014015}.

\bibitem[{Wu et~al.(2006)Wu, Zhong, and Zuo}]{Wu:2006rd}
\bibinfo{author}{Y.-L. Wu}, \bibinfo{author}{M.~Zhong}, \bibinfo{author}{Y.-B.
  Zuo}, \bibinfo{title}{{$B_{(s)}, D_{(s)} \to \pi, K, \eta, \rho, K^\ast,
  \omega, \phi$ Transition Form Factors and Decay Rates with Extraction of the
  CKM parameters $|V_{ub}|$, $|V_{cs}|$, $|V_{cd}|$}}, \bibinfo{journal}{Int.
  J. Mod. Phys. A} \bibinfo{volume}{21} (\bibinfo{year}{2006})
  \bibinfo{pages}{6125--6172}.

\bibitem[{Khodjamirian et~al.(2007)Khodjamirian, Mannel, and
  Offen}]{Khodjamirian:2006st}
\bibinfo{author}{A.~Khodjamirian}, \bibinfo{author}{T.~Mannel},
  \bibinfo{author}{N.~Offen}, \bibinfo{title}{{Form-factors from light-cone sum
  rules with B-meson distribution amplitudes}}, \bibinfo{journal}{Phys. Rev. D}
  \bibinfo{volume}{75} (\bibinfo{year}{2007}) \bibinfo{pages}{054013}.

\bibitem[{Ivanov et~al.(2006)Ivanov, Korner, and Santorelli}]{Ivanov:2006ni}
\bibinfo{author}{M.~A. Ivanov}, \bibinfo{author}{J.~G. Korner},
  \bibinfo{author}{P.~Santorelli}, \bibinfo{title}{{Exclusive semileptonic and
  nonleptonic decays of the $B_c$ meson}}, \bibinfo{journal}{Phys. Rev. D}
  \bibinfo{volume}{73} (\bibinfo{year}{2006}) \bibinfo{pages}{054024}.

\bibitem[{L{\"u} et~al.(2007)L{\"u}, Wang, and Wei}]{Lu:2007sg}
\bibinfo{author}{C.-D. L{\"u}}, \bibinfo{author}{W.~Wang},
  \bibinfo{author}{Z.-T. Wei}, \bibinfo{title}{{Heavy-to-light form factors on
  the light cone}}, \bibinfo{journal}{Phys. Rev. D} \bibinfo{volume}{76}
  (\bibinfo{year}{2007}) \bibinfo{pages}{014013}.

\bibitem[{Ivanov et~al.(2007)Ivanov, K{\"o}rner, Kovalenko, and
  Roberts}]{Ivanov:2007cw}
\bibinfo{author}{M.~A. Ivanov}, \bibinfo{author}{J.~G. K{\"o}rner},
  \bibinfo{author}{S.~G. Kovalenko}, \bibinfo{author}{C.~D. Roberts},
  \bibinfo{title}{{B- to light-meson transition form factors}},
  \bibinfo{journal}{Phys. Rev. D} \bibinfo{volume}{76} (\bibinfo{year}{2007})
  \bibinfo{pages}{034018}.

\bibitem[{Barik et~al.(2009)Barik, Naimuddin, Dash, and Kar}]{Barik:2009zz}
\bibinfo{author}{N.~Barik}, \bibinfo{author}{S.~Naimuddin},
  \bibinfo{author}{P.~C. Dash}, \bibinfo{author}{S.~Kar},
  \bibinfo{title}{{Semileptonic decays of the Bc meson}},
  \bibinfo{journal}{Phys. Rev. D} \bibinfo{volume}{80} (\bibinfo{year}{2009})
  \bibinfo{pages}{074005}.

\bibitem[{Faustov and Galkin(2013)}]{Faustov:2013ima}
\bibinfo{author}{R.~N. Faustov}, \bibinfo{author}{V.~O. Galkin},
  \bibinfo{title}{{Charmless weak $B_s$ decays in the relativistic quark
  model}}, \bibinfo{journal}{Phys. Rev. D}
  \bibinfo{volume}{87}~(\bibinfo{number}{9}) (\bibinfo{year}{2013})
  \bibinfo{pages}{094028}.

\bibitem[{Faustov and Galkin(2014)}]{Faustov:2014bxa}
\bibinfo{author}{R.~N. Faustov}, \bibinfo{author}{V.~O. Galkin},
  \bibinfo{title}{{Weak Decays of $B_s$ Mesons}}, in:
  \bibinfo{booktitle}{{Helmholtz International Summer School on Physics of
  Heavy Quarks and Hadrons}}, \bibinfo{pages}{62--72}, \bibinfo{year}{2014}.

\bibitem[{Xiao et~al.(2014)Xiao, Fan, Wang, and Cheng}]{Xiao:2014ana}
\bibinfo{author}{Z.-J. Xiao}, \bibinfo{author}{Y.-Y. Fan},
  \bibinfo{author}{W.-F. Wang}, \bibinfo{author}{S.~Cheng},
  \bibinfo{title}{{The semileptonic decays of $B/B_s$ meson in the perturbative
  QCD approach: A short review}}, \bibinfo{journal}{Chin. Sci. Bull.}
  \bibinfo{volume}{59} (\bibinfo{year}{2014}) \bibinfo{pages}{3787--3800}.

\bibitem[{Wang and Shen(2015)}]{Wang:2015vgv}
\bibinfo{author}{Y.-M. Wang}, \bibinfo{author}{Y.-L. Shen},
  \bibinfo{title}{{QCD corrections to $B\to \pi$ form factors from light-cone
  sum rules}}, \bibinfo{journal}{Nucl. Phys. B} \bibinfo{volume}{898}
  (\bibinfo{year}{2015}) \bibinfo{pages}{563--604}.

\bibitem[{Shi et~al.(2016)Shi, Wang, and Zhao}]{Shi:2016gqt}
\bibinfo{author}{Y.-J. Shi}, \bibinfo{author}{W.~Wang}, \bibinfo{author}{Z.-X.
  Zhao}, \bibinfo{title}{{$B_c\rightarrow B_{sJ}$ form factors and $B_c$ decays
  into $B_{sJ}$ in covariant light-front approach}}, \bibinfo{journal}{Eur.
  Phys. J. C} \bibinfo{volume}{76}~(\bibinfo{number}{10})
  (\bibinfo{year}{2016}) \bibinfo{pages}{555}.

\bibitem[{L\"u et~al.(2019)L\"u, Shen, Wang, and Wei}]{Lu:2018cfc}
\bibinfo{author}{C.-D. L\"u}, \bibinfo{author}{Y.-L. Shen},
  \bibinfo{author}{Y.-M. Wang}, \bibinfo{author}{Y.-B. Wei},
  \bibinfo{title}{{QCD calculations of $B \to \pi, K$ form factors with
  higher-twist corrections}}, \bibinfo{journal}{JHEP} \bibinfo{volume}{01}
  (\bibinfo{year}{2019}) \bibinfo{pages}{024}.

\bibitem[{Gubernari et~al.(2019)Gubernari, Kokulu, and van
  Dyk}]{Gubernari:2018wyi}
\bibinfo{author}{N.~Gubernari}, \bibinfo{author}{A.~Kokulu},
  \bibinfo{author}{D.~van Dyk}, \bibinfo{title}{{$B\to P$ and $B\to V$ Form
  Factors from $B$-Meson Light-Cone Sum Rules beyond Leading Twist}},
  \bibinfo{journal}{JHEP} \bibinfo{volume}{01} (\bibinfo{year}{2019})
  \bibinfo{pages}{150}.

\bibitem[{Ivanov et~al.(2019)Ivanov, K{\"o}rner, Pandya, Santorelli, Soni, and
  Tran}]{Ivanov:2019nqd}
\bibinfo{author}{M.~A. Ivanov}, \bibinfo{author}{J.~G. K{\"o}rner},
  \bibinfo{author}{J.~N. Pandya}, \bibinfo{author}{P.~Santorelli},
  \bibinfo{author}{N.~R. Soni}, \bibinfo{author}{C.-T. Tran},
  \bibinfo{title}{{Exclusive semileptonic decays of D and D$_{s}$ mesons in the
  covariant confining quark model}}, \bibinfo{journal}{Front. Phys. (Beijing)}
  \bibinfo{volume}{14} (\bibinfo{year}{2019}) \bibinfo{pages}{64401}.

\bibitem[{Hu et~al.(2020)Hu, Jin, and Xiao}]{Hu:2019bdf}
\bibinfo{author}{X.-Q. Hu}, \bibinfo{author}{S.-P. Jin}, \bibinfo{author}{Z.-J.
  Xiao}, \bibinfo{title}{{Semileptonic decays $B/B_s \to (D^{(*)},D_s^{(*)}) l
  \nu_l$ in the PQCD approach with the lattice QCD input}},
  \bibinfo{journal}{Chin. Phys. C} \bibinfo{volume}{44}~(\bibinfo{number}{5})
  (\bibinfo{year}{2020}) \bibinfo{pages}{053102}.

\bibitem[{Zhang et~al.(2021{\natexlab{a}})Zhang, Kang, Guo, Dai, Luo, and
  Wang}]{Zhang:2020dla}
\bibinfo{author}{L.~Zhang}, \bibinfo{author}{X.-W. Kang},
  \bibinfo{author}{X.-H. Guo}, \bibinfo{author}{L.-Y. Dai},
  \bibinfo{author}{T.~Luo}, \bibinfo{author}{C.~Wang}, \bibinfo{title}{{A
  comprehensive study on the semileptonic decay of heavy flavor mesons}},
  \bibinfo{journal}{JHEP} \bibinfo{volume}{02}
  (\bibinfo{year}{2021}{\natexlab{a}}) \bibinfo{pages}{179}.

\bibitem[{Gonz\`alez-Sol\'\i{}s et~al.(2021)Gonz\`alez-Sol\'\i{}s, Masjuan, and
  Rojas}]{Gonzalez-Solis:2021pyh}
\bibinfo{author}{S.~Gonz\`alez-Sol\'\i{}s}, \bibinfo{author}{P.~Masjuan},
  \bibinfo{author}{C.~Rojas}, \bibinfo{title}{{Pad\'e approximants to $B
  \rightarrow \pi \ell \nu_{\ell}$ and $B_s \rightarrow \ell \nu_{\ell}$ and
  determination of $|V_{ub}|$}}, \bibinfo{journal}{Phys. Rev. D}
  \bibinfo{volume}{104}~(\bibinfo{number}{11}) (\bibinfo{year}{2021})
  \bibinfo{pages}{114041}.

\bibitem[{Na et~al.(2011)Na, Davies, Follana, Koponen, Lepage, and
  Shigemitsu}]{Na:2011mc}
\bibinfo{author}{H.~Na}, \bibinfo{author}{C.~T.~H. Davies},
  \bibinfo{author}{E.~Follana}, \bibinfo{author}{J.~Koponen},
  \bibinfo{author}{G.~P. Lepage}, \bibinfo{author}{J.~Shigemitsu},
  \bibinfo{title}{{$D \rightarrow \pi, l \nu$ Semileptonic Decays, $|V_{cd}|$
  and 2$^{nd}$ Row Unitarity from Lattice QCD}}, \bibinfo{journal}{Phys. Rev.
  D} \bibinfo{volume}{84} (\bibinfo{year}{2011}) \bibinfo{pages}{114505}.

\bibitem[{Bouchard et~al.(2014)Bouchard, Lepage, Monahan, Na, and
  Shigemitsu}]{Bouchard:2014ypa}
\bibinfo{author}{C.~M. Bouchard}, \bibinfo{author}{G.~P. Lepage},
  \bibinfo{author}{C.~Monahan}, \bibinfo{author}{H.~Na},
  \bibinfo{author}{J.~Shigemitsu}, \bibinfo{title}{{$B_s \to K \ell \nu$ form
  factors from lattice QCD}}, \bibinfo{journal}{Phys. Rev. D}
  \bibinfo{volume}{90} (\bibinfo{year}{2014}) \bibinfo{pages}{054506}.

\bibitem[{Flynn et~al.(2015)Flynn, Izubuchi, Kawanai, Lehner, Soni, Van~de
  Water, and Witzel}]{Flynn:2015mha}
\bibinfo{author}{J.~M. Flynn}, \bibinfo{author}{T.~Izubuchi},
  \bibinfo{author}{T.~Kawanai}, \bibinfo{author}{C.~Lehner},
  \bibinfo{author}{A.~Soni}, \bibinfo{author}{R.~S. Van~de Water},
  \bibinfo{author}{O.~Witzel}, \bibinfo{title}{{$B \to \pi \ell \nu$ and $B_s
  \to K \ell \nu$ form factors and $|V_{ub}|$ from 2+1-flavor lattice QCD with
  domain-wall light quarks and relativistic heavy quarks}},
  \bibinfo{journal}{Phys. Rev. D} \bibinfo{volume}{91}~(\bibinfo{number}{7})
  (\bibinfo{year}{2015}) \bibinfo{pages}{074510}.

\bibitem[{Bailey et~al.(2015)}]{FermilabLattice:2015mwy}
\bibinfo{author}{J.~A. Bailey}, et~al., \bibinfo{title}{{$|V_{ub}|$ from
  $B\to\pi\ell\nu$ decays and (2+1)-flavor lattice QCD}},
  \bibinfo{journal}{Phys. Rev. D} \bibinfo{volume}{92}~(\bibinfo{number}{1})
  (\bibinfo{year}{2015}) \bibinfo{pages}{014024}.

\bibitem[{Colquhoun et~al.(2016)Colquhoun, Davies, Koponen, Lytle, and
  McNeile}]{Colquhoun:2016osw}
\bibinfo{author}{B.~Colquhoun}, \bibinfo{author}{C.~Davies},
  \bibinfo{author}{J.~Koponen}, \bibinfo{author}{A.~Lytle},
  \bibinfo{author}{C.~McNeile}, \bibinfo{title}{{$B_c$ decays from highly
  improved staggered quarks and NRQCD}}, \bibinfo{journal}{PoS}
  \bibinfo{volume}{LATTICE2016} (\bibinfo{year}{2016}) \bibinfo{pages}{281}.

\bibitem[{Lubicz et~al.(2017{\natexlab{a}})Lubicz, Riggio, Salerno, Simula, and
  Tarantino}]{Lubicz:2017syv}
\bibinfo{author}{V.~Lubicz}, \bibinfo{author}{L.~Riggio},
  \bibinfo{author}{G.~Salerno}, \bibinfo{author}{S.~Simula},
  \bibinfo{author}{C.~Tarantino}, \bibinfo{title}{{Scalar and vector form
  factors of $D \to \pi(K) \ell \nu$ decays with $N_f=2+1+1$ twisted
  fermions}}, \bibinfo{journal}{Phys. Rev. D} \bibinfo{volume}{96}
  (\bibinfo{year}{2017}{\natexlab{a}}) \bibinfo{pages}{054514},
  \bibinfo{note}{[erratum: Phys. Rev.D \textbf{99}, 099902 (2019)]}.

\bibitem[{Monahan et~al.(2017)Monahan, Na, Bouchard, Lepage, and
  Shigemitsu}]{Monahan:2017uby}
\bibinfo{author}{C.~J. Monahan}, \bibinfo{author}{H.~Na},
  \bibinfo{author}{C.~M. Bouchard}, \bibinfo{author}{G.~P. Lepage},
  \bibinfo{author}{J.~Shigemitsu}, \bibinfo{title}{{$B_s \to D_s \ell \nu$ Form
  Factors and the Fragmentation Fraction Ratio $f_s/f_d$}},
  \bibinfo{journal}{Phys. Rev. D} \bibinfo{volume}{95}~(\bibinfo{number}{11})
  (\bibinfo{year}{2017}) \bibinfo{pages}{114506}.

\bibitem[{Bazavov et~al.(2019)}]{FermilabLattice:2019ikx}
\bibinfo{author}{A.~Bazavov}, et~al., \bibinfo{title}{{$B_s\to K\ell\nu$ decay
  from lattice QCD}}, \bibinfo{journal}{Phys. Rev. D}
  \bibinfo{volume}{100}~(\bibinfo{number}{3}) (\bibinfo{year}{2019})
  \bibinfo{pages}{034501}.

\bibitem[{McLean et~al.(2020)McLean, Davies, Koponen, and
  Lytle}]{McLean:2019qcx}
\bibinfo{author}{E.~McLean}, \bibinfo{author}{C.~T.~H. Davies},
  \bibinfo{author}{J.~Koponen}, \bibinfo{author}{A.~T. Lytle},
  \bibinfo{title}{{$B_s\to D_s \ell\nu$ Form Factors for the full $q^2$ range
  from Lattice QCD with non-perturbatively normalized currents}},
  \bibinfo{journal}{Phys. Rev. D} \bibinfo{volume}{101} (\bibinfo{year}{2020})
  \bibinfo{pages}{074513}.

\bibitem[{Cooper et~al.(2020)Cooper, Davies, Harrison, Komijani, and
  Wingate}]{Cooper:2020wnj}
\bibinfo{author}{L.~J. Cooper}, \bibinfo{author}{C.~T.~H. Davies},
  \bibinfo{author}{J.~Harrison}, \bibinfo{author}{J.~Komijani},
  \bibinfo{author}{M.~Wingate}, \bibinfo{title}{{$B_c \to B_{s(d)}$ form
  factors from lattice QCD}}, \bibinfo{journal}{Phys. Rev. D}
  \bibinfo{volume}{102}~(\bibinfo{number}{1}) (\bibinfo{year}{2020})
  \bibinfo{pages}{014513}, \bibinfo{note}{[Erratum: Phys.Rev.D 103, 099901
  (2021)]}.

\bibitem[{Chakraborty et~al.(2021)Chakraborty, Parrott, Bouchard, Davies,
  Koponen, and Lepage}]{Chakraborty:2021qav}
\bibinfo{author}{B.~Chakraborty}, \bibinfo{author}{W.~G. Parrott},
  \bibinfo{author}{C.~Bouchard}, \bibinfo{author}{C.~T.~H. Davies},
  \bibinfo{author}{J.~Koponen}, \bibinfo{author}{G.~P. Lepage},
  \bibinfo{title}{{Improved Vcs determination using precise lattice QCD form
  factors for D\textrightarrow{}K\ensuremath{\ell}\ensuremath{\nu}}},
  \bibinfo{journal}{Phys. Rev. D} \bibinfo{volume}{104}~(\bibinfo{number}{3})
  (\bibinfo{year}{2021}) \bibinfo{pages}{034505}.

\bibitem[{Yao et~al.(2022)Yao, Binosi, Cui, and Roberts}]{Yao:2021pdy}
\bibinfo{author}{Z.-Q. Yao}, \bibinfo{author}{D.~Binosi},
  \bibinfo{author}{Z.-F. Cui}, \bibinfo{author}{C.~D. Roberts},
  \bibinfo{title}{{Semileptonic transitions: $B_{(s)} \to \pi(K)$; $D_s \to K$;
  $D\to \pi, K$; and $K\to \pi$}}, \bibinfo{journal}{Phys. Lett. B}
  \bibinfo{volume}{824} (\bibinfo{year}{2022}) \bibinfo{pages}{136793}.

\bibitem[{Yao et~al.(2021)Yao, Binosi, Cui, and Roberts}]{Yao:2021pyf}
\bibinfo{author}{Z.-Q. Yao}, \bibinfo{author}{D.~Binosi},
  \bibinfo{author}{Z.-F. Cui}, \bibinfo{author}{C.~D. Roberts},
  \bibinfo{title}{{Semileptonic $B_c \to \eta_c, J/\psi$ transitions}},
  \bibinfo{journal}{Phys. Lett. B} \bibinfo{volume}{818} (\bibinfo{year}{2021})
  \bibinfo{pages}{136344}.

\bibitem[{Nayak et~al.(2021)Nayak, Patnaik, Dash, Kar, and
  Barik}]{Nayak:2021djn}
\bibinfo{author}{L.~Nayak}, \bibinfo{author}{S.~Patnaik},
  \bibinfo{author}{P.~C. Dash}, \bibinfo{author}{S.~Kar},
  \bibinfo{author}{N.~Barik}, \bibinfo{title}{{Lepton mass effects in exclusive
  semileptonic $B_c$-meson decays}}, \bibinfo{journal}{Phys. Rev. D}
  \bibinfo{volume}{104} (\bibinfo{year}{2021}) \bibinfo{pages}{036012}.

\bibitem[{Gao et~al.(2022)Gao, Huber, Ji, Wang, Wang, and Wei}]{Gao:2021sav}
\bibinfo{author}{J.~Gao}, \bibinfo{author}{T.~Huber}, \bibinfo{author}{Y.~Ji},
  \bibinfo{author}{C.~Wang}, \bibinfo{author}{Y.-M. Wang},
  \bibinfo{author}{Y.-B. Wei}, \bibinfo{title}{{$B \to D \ell \nu_\ell$ form
  factors beyond leading power and extraction of |V$_{cb}$| and R(D)}},
  \bibinfo{journal}{JHEP} \bibinfo{volume}{05} (\bibinfo{year}{2022})
  \bibinfo{pages}{024}.

\bibitem[{Ball and Zwicky(2005{\natexlab{b}})}]{Ball:2004rg}
\bibinfo{author}{P.~Ball}, \bibinfo{author}{R.~Zwicky},
  \bibinfo{title}{{$B_{d,s} \to \rho, \omega, K^*, \phi$ decay form-factors
  from light-cone sum rules revisited}}, \bibinfo{journal}{Phys. Rev. D}
  \bibinfo{volume}{71} (\bibinfo{year}{2005}{\natexlab{b}})
  \bibinfo{pages}{014029}.

\bibitem[{Donald et~al.(2014)Donald, Davies, Koponen, and
  Lepage}]{Donald:2013pea}
\bibinfo{author}{G.~C. Donald}, \bibinfo{author}{C.~T.~H. Davies},
  \bibinfo{author}{J.~Koponen}, \bibinfo{author}{G.~P. Lepage},
  \bibinfo{title}{{$V_{cs}$ from $D_s \to \phi \ell \nu$ semileptonic decay and
  full lattice QCD}}, \bibinfo{journal}{Phys. Rev. D}
  \bibinfo{volume}{90}~(\bibinfo{number}{7}) (\bibinfo{year}{2014})
  \bibinfo{pages}{074506}.

\bibitem[{Sekihara and Oset(2015)}]{Sekihara:2015iha}
\bibinfo{author}{T.~Sekihara}, \bibinfo{author}{E.~Oset},
  \bibinfo{title}{{Investigating the nature of light scalar mesons with
  semileptonic decays of D mesons}}, \bibinfo{journal}{Phys. Rev. D}
  \bibinfo{volume}{92}~(\bibinfo{number}{5}) (\bibinfo{year}{2015})
  \bibinfo{pages}{054038}.

\bibitem[{Chang et~al.(2019)Chang, Li, and Wang}]{Chang:2019mmh}
\bibinfo{author}{Q.~Chang}, \bibinfo{author}{X.-N. Li}, \bibinfo{author}{L.-T.
  Wang}, \bibinfo{title}{{Revisiting the form factors of $P\rightarrow V$
  transition within the light-front quark models}}, \bibinfo{journal}{Eur.
  Phys. J. C} \bibinfo{volume}{79}~(\bibinfo{number}{5}) (\bibinfo{year}{2019})
  \bibinfo{pages}{422}.

\bibitem[{Roberts et~al.(2010)Roberts, Roberts, Bashir, Guti{\'e}rrez-Guerrero,
  and Tandy}]{Roberts:2010rn}
\bibinfo{author}{H.~L.~L. Roberts}, \bibinfo{author}{C.~D. Roberts},
  \bibinfo{author}{A.~Bashir}, \bibinfo{author}{L.~X. Guti{\'e}rrez-Guerrero},
  \bibinfo{author}{P.~C. Tandy}, \bibinfo{title}{{Abelian anomaly and neutral
  pion production}}, \bibinfo{journal}{Phys. Rev. C} \bibinfo{volume}{82}
  (\bibinfo{year}{2010}) \bibinfo{pages}{{\mbox{065202}}}.

\bibitem[{Roberts et~al.(2011)Roberts, Bashir, Guti{\'e}rrez-Guerrero, Roberts,
  and Wilson}]{Roberts:2011wy}
\bibinfo{author}{H.~L.~L. Roberts}, \bibinfo{author}{A.~Bashir},
  \bibinfo{author}{L.~X. Guti{\'e}rrez-Guerrero}, \bibinfo{author}{C.~D.
  Roberts}, \bibinfo{author}{D.~J. Wilson}, \bibinfo{title}{{$\pi$- and
  $\rho$-mesons, and their diquark partners, from a contact interaction}},
  \bibinfo{journal}{Phys. Rev. C} \bibinfo{volume}{83} (\bibinfo{year}{2011})
  \bibinfo{pages}{065206}.

\bibitem[{Wang et~al.(2013)Wang, Liu, Chang, Roberts, and
  Schmidt}]{Wang:2013wk}
\bibinfo{author}{K.-L. Wang}, \bibinfo{author}{Y.-X. Liu},
  \bibinfo{author}{L.~Chang}, \bibinfo{author}{C.~D. Roberts},
  \bibinfo{author}{S.~M. Schmidt}, \bibinfo{title}{{Baryon and meson screening
  masses}}, \bibinfo{journal}{Phys. Rev. D} \bibinfo{volume}{87}
  (\bibinfo{year}{2013}) \bibinfo{pages}{074038}.

\bibitem[{Segovia et~al.(2014)Segovia, Cloet, Roberts, and
  Schmidt}]{Segovia:2014aza}
\bibinfo{author}{J.~Segovia}, \bibinfo{author}{I.~C. Cloet},
  \bibinfo{author}{C.~D. Roberts}, \bibinfo{author}{S.~M. Schmidt},
  \bibinfo{title}{{Nucleon and $\Delta$ elastic and transition form factors}},
  \bibinfo{journal}{Few Body Syst.} \bibinfo{volume}{55} (\bibinfo{year}{2014})
  \bibinfo{pages}{1185--1222}.

\bibitem[{Xu et~al.(2015)Xu, Chen, Cloet, Roberts, Segovia, and
  Zong}]{Xu:2015kta}
\bibinfo{author}{S.-S. Xu}, \bibinfo{author}{C.~Chen}, \bibinfo{author}{I.~C.
  Cloet}, \bibinfo{author}{C.~D. Roberts}, \bibinfo{author}{J.~Segovia},
  \bibinfo{author}{H.-S. Zong}, \bibinfo{title}{{Contact-interaction Faddeev
  equation and, \emph{inter alia}, proton tensor charges}},
  \bibinfo{journal}{Phys. Rev. D} \bibinfo{volume}{92} (\bibinfo{year}{2015})
  \bibinfo{pages}{114034}.

\bibitem[{Bedolla et~al.(2015)Bedolla, Cobos-Mart{\'\i}nez, and
  Bashir}]{Bedolla:2015mpa}
\bibinfo{author}{M.~A. Bedolla}, \bibinfo{author}{J.~J. Cobos-Mart{\'\i}nez},
  \bibinfo{author}{A.~Bashir}, \bibinfo{title}{{Charmonia in a contact
  interaction}}, \bibinfo{journal}{Phys. Rev. D} \bibinfo{volume}{92}
  (\bibinfo{year}{2015}) \bibinfo{pages}{054031}.

\bibitem[{Bedolla et~al.(2016)Bedolla, Raya, Cobos-Mart{\'\i}nez, and
  Bashir}]{Bedolla:2016yxq}
\bibinfo{author}{M.~A. Bedolla}, \bibinfo{author}{K.~Raya},
  \bibinfo{author}{J.~J. Cobos-Mart{\'\i}nez}, \bibinfo{author}{A.~Bashir},
  \bibinfo{title}{{\mbox{$\eta_c$} elastic and transition form factors: Contact
  interaction and algebraic model}}, \bibinfo{journal}{Phys. Rev. D}
  \bibinfo{volume}{93} (\bibinfo{year}{2016}) \bibinfo{pages}{094025}.

\bibitem[{Serna et~al.(2017)Serna, El-Bennich, and Krein}]{Serna:2017nlr}
\bibinfo{author}{F.~E. Serna}, \bibinfo{author}{B.~El-Bennich},
  \bibinfo{author}{G.~Krein}, \bibinfo{title}{{Charmed mesons with a
  symmetry-preserving contact interaction}}, \bibinfo{journal}{Phys. Rev. D}
  \bibinfo{volume}{96} (\bibinfo{year}{2017}) \bibinfo{pages}{014013}.

\bibitem[{Raya et~al.(2018)Raya, Bedolla, Cobos-Mart{\'{\i}}nez, and
  Bashir}]{Raya:2017ggu}
\bibinfo{author}{K.~Raya}, \bibinfo{author}{M.~A. Bedolla},
  \bibinfo{author}{J.~J. Cobos-Mart{\'{\i}}nez}, \bibinfo{author}{A.~Bashir},
  \bibinfo{title}{{Heavy quarkonia in a contact interaction and an algebraic
  model: mass spectrum, decay constants, charge radii and elastic and
  transition form factors}}, \bibinfo{journal}{Few Body Syst.}
  \bibinfo{volume}{59} (\bibinfo{year}{2018}) \bibinfo{pages}{133}.

\bibitem[{Zhang et~al.(2021{\natexlab{b}})Zhang, Cui, Ping, and
  Roberts}]{Zhang:2020ecj}
\bibinfo{author}{J.-L. Zhang}, \bibinfo{author}{Z.-F. Cui},
  \bibinfo{author}{J.~Ping}, \bibinfo{author}{C.~D. Roberts},
  \bibinfo{title}{{Contact interaction analysis of pion GTMDs}},
  \bibinfo{journal}{Eur. Phys. J. C} \bibinfo{volume}{81}~(\bibinfo{number}{1})
  (\bibinfo{year}{2021}{\natexlab{b}}) \bibinfo{pages}{6}.

\bibitem[{Yin et~al.(2021)Yin, Cui, Roberts, and Segovia}]{Yin:2021uom}
\bibinfo{author}{P.-L. Yin}, \bibinfo{author}{Z.-F. Cui},
  \bibinfo{author}{C.~D. Roberts}, \bibinfo{author}{J.~Segovia},
  \bibinfo{title}{{Masses of positive- and negative-parity hadron
  ground-states, including those with heavy quarks}}, \bibinfo{journal}{Eur.
  Phys. J. C} \bibinfo{volume}{81}~(\bibinfo{number}{4}) (\bibinfo{year}{2021})
  \bibinfo{pages}{327}.

\bibitem[{Raya et~al.(2021)Raya, Guti\'errez-Guerrero, Bashir, Chang, Cui, Lu,
  Roberts, and Segovia}]{Raya:2021pyr}
\bibinfo{author}{K.~Raya}, \bibinfo{author}{L.~X. Guti\'errez-Guerrero},
  \bibinfo{author}{A.~Bashir}, \bibinfo{author}{L.~Chang},
  \bibinfo{author}{Z.~F. Cui}, \bibinfo{author}{Y.~Lu}, \bibinfo{author}{C.~D.
  Roberts}, \bibinfo{author}{J.~Segovia}, \bibinfo{title}{{Dynamical diquarks
  in the \mbox{$\gamma^{(\ast)} p\to N(1535)\tfrac{1}{2}^-$} transition}},
  \bibinfo{journal}{Eur. Phys. J. A} \bibinfo{volume}{57}~(\bibinfo{number}{9})
  (\bibinfo{year}{2021}) \bibinfo{pages}{266}.

\bibitem[{Lu et~al.(2021)Lu, Binosi, Ding, Roberts, Xing, and Xu}]{Lu:2021sgg}
\bibinfo{author}{Y.~Lu}, \bibinfo{author}{D.~Binosi},
  \bibinfo{author}{M.~Ding}, \bibinfo{author}{C.~D. Roberts},
  \bibinfo{author}{H.-Y. Xing}, \bibinfo{author}{C.~Xu},
  \bibinfo{title}{{Distribution amplitudes of light diquarks}},
  \bibinfo{journal}{Eur. Phys. J A (Lett)}
  \bibinfo{volume}{57}~(\bibinfo{number}{4}) (\bibinfo{year}{2021})
  \bibinfo{pages}{115}.

\bibitem[{Guti\'errez-Guerrero et~al.(2021)Guti\'errez-Guerrero,
  Paredes-Torres, and Bashir}]{Gutierrez-Guerrero:2021rsx}
\bibinfo{author}{L.~X. Guti\'errez-Guerrero},
  \bibinfo{author}{G.~Paredes-Torres}, \bibinfo{author}{A.~Bashir},
  \bibinfo{title}{{Mesons and baryons: Parity partners}},
  \bibinfo{journal}{Phys. Rev. D} \bibinfo{volume}{104}~(\bibinfo{number}{9})
  (\bibinfo{year}{2021}) \bibinfo{pages}{094013}.

\bibitem[{Zyla et~al.(2020)}]{Zyla:2020zbs}
\bibinfo{author}{P.~Zyla}, et~al., \bibinfo{title}{{Review of Particle
  Physics}}, \bibinfo{journal}{PTEP} \bibinfo{volume}{2020}
  (\bibinfo{year}{2020}) \bibinfo{pages}{083C01}.

\bibitem[{Qin and Roberts(2020)}]{Qin:2020rad}
\bibinfo{author}{S.-X. Qin}, \bibinfo{author}{C.~D. Roberts},
  \bibinfo{title}{{Impressions of the Continuum Bound State Problem in QCD}},
  \bibinfo{journal}{Chin. Phys. Lett.}
  \bibinfo{volume}{37}~(\bibinfo{number}{12}) (\bibinfo{year}{2020})
  \bibinfo{pages}{121201}.

\bibitem[{Aaij et~al.(2018)}]{Aaij:2017tyk}
\bibinfo{author}{R.~Aaij}, et~al., \bibinfo{title}{{Measurement of the ratio of
  branching fractions
  $\mathcal{B}(B_c^+\,\to\,J/\psi\tau^+\nu_\tau)$/$\mathcal{B}(B_c^+\,\to\,J/\psi\mu^+\nu_\mu)$}},
  \bibinfo{journal}{Phys. Rev. Lett.}
  \bibinfo{volume}{120}~(\bibinfo{number}{12}) (\bibinfo{year}{2018})
  \bibinfo{pages}{121801}.

\bibitem[{Dobbs et~al.(2013)}]{CLEO:2011ab}
\bibinfo{author}{S.~Dobbs}, et~al., \bibinfo{title}{{First Measurement of the
  Form Factors in the Decays $D^0 \to \rho^- e^+ \nu_e$ and $D^+ \to \rho^0 e^+
  \nu_e$}}, \bibinfo{journal}{Phys. Rev. Lett.}
  \bibinfo{volume}{110}~(\bibinfo{number}{13}) (\bibinfo{year}{2013})
  \bibinfo{pages}{131802}.

\bibitem[{Ablikim et~al.(2019{\natexlab{a}})}]{BESIII:2018qmf}
\bibinfo{author}{M.~Ablikim}, et~al., \bibinfo{title}{{Observation of $D^+ \to
  f_0(500) e^+\nu_e$ and Improved Measurements of $D \to\rho e^+\nu_e$}},
  \bibinfo{journal}{Phys. Rev. Lett.}
  \bibinfo{volume}{122}~(\bibinfo{number}{6})
  (\bibinfo{year}{2019}{\natexlab{a}}) \bibinfo{pages}{062001}.

\bibitem[{Nakamura et~al.(2010)}]{Nakamura:2010zzi}
\bibinfo{author}{K.~Nakamura}, et~al., \bibinfo{title}{{Review of particle
  physics}}, \bibinfo{journal}{J. Phys. G} \bibinfo{volume}{37}
  (\bibinfo{year}{2010}) \bibinfo{pages}{075021}.

\bibitem[{Yelton et~al.(2009)}]{CLEO:2009dyb}
\bibinfo{author}{J.~Yelton}, et~al., \bibinfo{title}{{Absolute Branching
  Fraction Measurements for Exclusive D(s) Semileptonic Decays}},
  \bibinfo{journal}{Phys. Rev. D} \bibinfo{volume}{80} (\bibinfo{year}{2009})
  \bibinfo{pages}{052007}.

\bibitem[{Hietala et~al.(2015)Hietala, Cronin-Hennessy, Pedlar, and
  Shipsey}]{Hietala:2015jqa}
\bibinfo{author}{J.~Hietala}, \bibinfo{author}{D.~Cronin-Hennessy},
  \bibinfo{author}{T.~Pedlar}, \bibinfo{author}{I.~Shipsey},
  \bibinfo{title}{{Exclusive $D_s$ semileptonic branching fraction
  measurements}}, \bibinfo{journal}{Phys. Rev. D}
  \bibinfo{volume}{92}~(\bibinfo{number}{1}) (\bibinfo{year}{2015})
  \bibinfo{pages}{012009}.

\bibitem[{Ablikim et~al.(2019{\natexlab{b}})}]{BESIII:2018xre}
\bibinfo{author}{M.~Ablikim}, et~al., \bibinfo{title}{{First Measurement of the
  Form Factors in $D^+_{s}\rightarrow K^0 e^+\nu_e$ and $D^+_{s}\rightarrow
  K^{*0} e^+\nu_e$ Decays}}, \bibinfo{journal}{Phys. Rev. Lett.}
  \bibinfo{volume}{122}~(\bibinfo{number}{6})
  (\bibinfo{year}{2019}{\natexlab{b}}) \bibinfo{pages}{061801}.

\bibitem[{Link et~al.(2005)}]{FOCUS:2004zbs}
\bibinfo{author}{J.~M. Link}, et~al., \bibinfo{title}{{Analysis of the
  semileptonic decay D0 ---\ensuremath{>} anti-K0 pi- mu+ nu}},
  \bibinfo{journal}{Phys. Lett. B} \bibinfo{volume}{607} (\bibinfo{year}{2005})
  \bibinfo{pages}{67--77}.

\bibitem[{Ablikim et~al.(2019{\natexlab{c}})}]{BESIII:2018jjm}
\bibinfo{author}{M.~Ablikim}, et~al., \bibinfo{title}{{Study of the decay
  $D^0\rightarrow \bar{K}^0\pi^-e^+\nu_e$}}, \bibinfo{journal}{Phys. Rev. D}
  \bibinfo{volume}{99}~(\bibinfo{number}{1})
  (\bibinfo{year}{2019}{\natexlab{c}}) \bibinfo{pages}{011103}.

\bibitem[{Bowler et~al.(2004)Bowler, Gill, Maynard, and Flynn}]{Bowler:2004zb}
\bibinfo{author}{K.~C. Bowler}, \bibinfo{author}{J.~F. Gill},
  \bibinfo{author}{C.~M. Maynard}, \bibinfo{author}{J.~M. Flynn},
  \bibinfo{title}{{$B \to \rho \ell \nu$ form-factors in lattice QCD}},
  \bibinfo{journal}{JHEP} \bibinfo{volume}{05} (\bibinfo{year}{2004})
  \bibinfo{pages}{035}.

\bibitem[{Horgan et~al.(2014)Horgan, Liu, Meinel, and Wingate}]{Horgan:2013hoa}
\bibinfo{author}{R.~R. Horgan}, \bibinfo{author}{Z.~Liu},
  \bibinfo{author}{S.~Meinel}, \bibinfo{author}{M.~Wingate},
  \bibinfo{title}{{Lattice QCD calculation of form factors describing the rare
  decays $B \to K^* \ell^+ \ell^-$ and $B_s \to \phi \ell^+ \ell^-$}},
  \bibinfo{journal}{Phys. Rev. D} \bibinfo{volume}{89}~(\bibinfo{number}{9})
  (\bibinfo{year}{2014}) \bibinfo{pages}{094501}.

\bibitem[{Fajfer et~al.(2012)Fajfer, Kamenik, and Nisandzic}]{Fajfer:2012vx}
\bibinfo{author}{S.~Fajfer}, \bibinfo{author}{J.~F. Kamenik},
  \bibinfo{author}{I.~Nisandzic}, \bibinfo{title}{{On the $B \to D^* \tau \bar
  \nu_{\tau}$ Sensitivity to New Physics}}, \bibinfo{journal}{Phys. Rev. D}
  \bibinfo{volume}{85} (\bibinfo{year}{2012}) \bibinfo{pages}{094025}.

\bibitem[{Amhis et~al.(2021)}]{HFLAV:2019otj}
\bibinfo{author}{Y.~S. Amhis}, et~al., \bibinfo{title}{{Averages of b-hadron,
  c-hadron, and $\tau $-lepton properties as of 2018}}, \bibinfo{journal}{Eur.
  Phys. J. C} \bibinfo{volume}{81}~(\bibinfo{number}{3}) (\bibinfo{year}{2021})
  \bibinfo{pages}{226}.

\bibitem[{Amhis et~al.(2017)}]{HFLAV:2016hnz}
\bibinfo{author}{Y.~Amhis}, et~al., \bibinfo{title}{{Averages of $b$-hadron,
  $c$-hadron, and $\tau$-lepton properties as of summer 2016}},
  \bibinfo{journal}{Eur. Phys. J. C}
  \bibinfo{volume}{77}~(\bibinfo{number}{12}) (\bibinfo{year}{2017})
  \bibinfo{pages}{895}.

\bibitem[{Aaij et~al.(2020)}]{LHCb:2020cyw}
\bibinfo{author}{R.~Aaij}, et~al., \bibinfo{title}{{Measurement of $|V_{cb}|$
  with $B_s^0 \to D_s^{(*)-} \mu^+ \nu_{\mu}$ decays}}, \bibinfo{journal}{Phys.
  Rev. D} \bibinfo{volume}{101}~(\bibinfo{number}{7}) (\bibinfo{year}{2020})
  \bibinfo{pages}{072004}.

\bibitem[{Harrison and Davies(2021)}]{Harrison:2021tol}
\bibinfo{author}{J.~Harrison}, \bibinfo{author}{C.~T.~H. Davies},
  \bibinfo{title}{{$B_s \rightarrow D_s^*$ Form Factors for the full $q^2$
  range from Lattice QCD -- arXiv:2105.11433 [hep-lat]}} .

\bibitem[{Isgur and Wise(1990)}]{Isgur:1989ed}
\bibinfo{author}{N.~Isgur}, \bibinfo{author}{M.~B. Wise}, \bibinfo{title}{{Weak
  Transition Form Factors between Heavy Mesons}}, \bibinfo{journal}{Phys. Lett.
  B} \bibinfo{volume}{237} (\bibinfo{year}{1990}) \bibinfo{pages}{527--530}.

\bibitem[{McLean et~al.(2019)McLean, Davies, Lytle, and
  Koponen}]{McLean:2019sds}
\bibinfo{author}{E.~McLean}, \bibinfo{author}{C.~T.~H. Davies},
  \bibinfo{author}{A.~T. Lytle}, \bibinfo{author}{J.~Koponen},
  \bibinfo{title}{{Lattice QCD form factor for $B_s\to D_s^* l\nu$ at zero
  recoil with non-perturbative current renormalisation}},
  \bibinfo{journal}{Phys. Rev. D} \bibinfo{volume}{99}~(\bibinfo{number}{11})
  (\bibinfo{year}{2019}) \bibinfo{pages}{114512}.

\bibitem[{Glattauer et~al.(2016)}]{Glattauer:2015teq}
\bibinfo{author}{R.~Glattauer}, et~al., \bibinfo{title}{{Measurement of the
  decay $B\to D\ell\nu_\ell$ in fully reconstructed events and determination of
  the Cabibbo-Kobayashi-Maskawa matrix element $|V_{cb}|$}},
  \bibinfo{journal}{Phys. Rev. D} \bibinfo{volume}{93}~(\bibinfo{number}{3})
  (\bibinfo{year}{2016}) \bibinfo{pages}{032006}.

\bibitem[{Harrison et~al.(2020)Harrison, Davies, and Lytle}]{Harrison:2020gvo}
\bibinfo{author}{J.~Harrison}, \bibinfo{author}{C.~T.~H. Davies},
  \bibinfo{author}{A.~Lytle}, \bibinfo{title}{{$B_c \rightarrow J/\psi$ form
  factors for the full $q^2$ range from lattice QCD}}, \bibinfo{journal}{Phys.
  Rev. D} \bibinfo{volume}{102} (\bibinfo{year}{2020}) \bibinfo{pages}{094518}.

\bibitem[{Boucaud et~al.(2012)Boucaud, Leroy, Le-Yaouanc, Micheli, Pene, and
  Rodr{\'i}guez-Quintero}]{Boucaud:2011ug}
\bibinfo{author}{P.~Boucaud}, \bibinfo{author}{J.~P. Leroy},
  \bibinfo{author}{A.~Le-Yaouanc}, \bibinfo{author}{J.~Micheli},
  \bibinfo{author}{O.~Pene}, \bibinfo{author}{J.~Rodr{\'i}guez-Quintero},
  \bibinfo{title}{{The Infrared Behaviour of the Pure Yang-Mills Green
  Functions}}, \bibinfo{journal}{Few Body Syst.} \bibinfo{volume}{53}
  (\bibinfo{year}{2012}) \bibinfo{pages}{387--436}.

\bibitem[{Aguilar et~al.(2016)Aguilar, Binosi, and
  Papavassiliou}]{Aguilar:2015bud}
\bibinfo{author}{A.~C. Aguilar}, \bibinfo{author}{D.~Binosi},
  \bibinfo{author}{J.~Papavassiliou}, \bibinfo{title}{{The Gluon Mass
  Generation Mechanism: A Concise Primer}}, \bibinfo{journal}{Front. Phys.
  China} \bibinfo{volume}{11} (\bibinfo{year}{2016}) \bibinfo{pages}{111203}.

\bibitem[{Cui et~al.(2020)Cui, Zhang, Binosi, de~Soto, Mezrag, Papavassiliou,
  Roberts, Rodr{\'{\i}}guez-Quintero, Segovia, and Zafeiropoulos}]{Cui:2019dwv}
\bibinfo{author}{Z.-F. Cui}, \bibinfo{author}{J.-L. Zhang},
  \bibinfo{author}{D.~Binosi}, \bibinfo{author}{F.~de~Soto},
  \bibinfo{author}{C.~Mezrag}, \bibinfo{author}{J.~Papavassiliou},
  \bibinfo{author}{C.~D. Roberts},
  \bibinfo{author}{J.~Rodr{\'{\i}}guez-Quintero}, \bibinfo{author}{J.~Segovia},
  \bibinfo{author}{S.~Zafeiropoulos}, \bibinfo{title}{{Effective charge from
  lattice QCD}}, \bibinfo{journal}{Chin. Phys. C} \bibinfo{volume}{44}
  (\bibinfo{year}{2020}) \bibinfo{pages}{083102}.

\bibitem[{Ebert et~al.(1996)Ebert, Feldmann, and Reinhardt}]{Ebert:1996vx}
\bibinfo{author}{D.~Ebert}, \bibinfo{author}{T.~Feldmann},
  \bibinfo{author}{H.~Reinhardt}, \bibinfo{title}{{Extended NJL model for light
  and heavy mesons without $q \bar q$ thresholds}}, \bibinfo{journal}{Phys.
  Lett. B} \bibinfo{volume}{388} (\bibinfo{year}{1996})
  \bibinfo{pages}{154--160}.

\bibitem[{Roberts et~al.(1992)Roberts, Williams, and Krein}]{Krein:1990sf}
\bibinfo{author}{C.~D. Roberts}, \bibinfo{author}{A.~G. Williams},
  \bibinfo{author}{G.~Krein}, \bibinfo{title}{{On the implications of
  confinement}}, \bibinfo{journal}{Int. J. Mod. Phys. A} \bibinfo{volume}{7}
  (\bibinfo{year}{1992}) \bibinfo{pages}{5607--5624}.

\bibitem[{Cui et~al.(2022)Cui, Binosi, Roberts, and Schmidt}]{Cui:2022fyr}
\bibinfo{author}{Z.-F. Cui}, \bibinfo{author}{D.~Binosi},
  \bibinfo{author}{C.~D. Roberts}, \bibinfo{author}{S.~M. Schmidt},
  \bibinfo{title}{{Hadron and light nucleus radii from electron scattering --
  arXiv:2204.05418 [hep-ph]}} .

\bibitem[{Guti{\'e}rrez-Guerrero et~al.(2019)Guti{\'e}rrez-Guerrero, Bashir,
  Bedolla, and Santopinto}]{Gutierrez-Guerrero:2019uwa}
\bibinfo{author}{L.~Guti{\'e}rrez-Guerrero}, \bibinfo{author}{A.~Bashir},
  \bibinfo{author}{M.~A. Bedolla}, \bibinfo{author}{E.~Santopinto},
  \bibinfo{title}{{Masses of Light and Heavy Mesons and Baryons: A Unified
  Picture}}, \bibinfo{journal}{Phys. Rev. D} \bibinfo{volume}{100}
  (\bibinfo{year}{2019}) \bibinfo{pages}{114032}.

\bibitem[{Guti{\'e}rrez-Guerrero et~al.(2010)Guti{\'e}rrez-Guerrero, Bashir,
  Cloet, and Roberts}]{GutierrezGuerrero:2010md}
\bibinfo{author}{L.~X. Guti{\'e}rrez-Guerrero}, \bibinfo{author}{A.~Bashir},
  \bibinfo{author}{I.~C. Cloet}, \bibinfo{author}{C.~D. Roberts},
  \bibinfo{title}{{Pion form factor from a contact interaction}},
  \bibinfo{journal}{Phys. Rev. C} \bibinfo{volume}{81} (\bibinfo{year}{2010})
  \bibinfo{pages}{065202}.

\bibitem[{Mathur et~al.(2018)Mathur, Padmanath, and Mondal}]{Mathur:2018epb}
\bibinfo{author}{N.~Mathur}, \bibinfo{author}{M.~Padmanath},
  \bibinfo{author}{S.~Mondal}, \bibinfo{title}{{Precise predictions of
  charmed-bottom hadrons from lattice QCD}}, \bibinfo{journal}{Phys. Rev.
  Lett.} \bibinfo{volume}{121}~(\bibinfo{number}{20}) (\bibinfo{year}{2018})
  \bibinfo{pages}{202002}.

\bibitem[{Aoki et~al.(2020)}]{Aoki:2019cca}
\bibinfo{author}{S.~Aoki}, et~al., \bibinfo{title}{{FLAG Review 2019}},
  \bibinfo{journal}{Eur. Phys. J. C} \bibinfo{volume}{80}
  (\bibinfo{year}{2020}) \bibinfo{pages}{113}.

\bibitem[{McNeile et~al.(2012)McNeile, Davies, Follana, Hornbostel, and
  Lepage}]{McNeile:2012qf}
\bibinfo{author}{C.~McNeile}, \bibinfo{author}{C.~T.~H. Davies},
  \bibinfo{author}{E.~Follana}, \bibinfo{author}{K.~Hornbostel},
  \bibinfo{author}{G.~P. Lepage}, \bibinfo{title}{{Heavy meson masses and decay
  constants from relativistic heavy quarks in full lattice QCD}},
  \bibinfo{journal}{Phys. Rev. D} \bibinfo{volume}{86} (\bibinfo{year}{2012})
  \bibinfo{pages}{074503}.

\bibitem[{Colquhoun et~al.(2015)Colquhoun, Davies, Dowdall, Kettle, Koponen,
  Lepage, and Lytle}]{Colquhoun:2015oha}
\bibinfo{author}{B.~Colquhoun}, \bibinfo{author}{C.~T.~H. Davies},
  \bibinfo{author}{R.~J. Dowdall}, \bibinfo{author}{J.~Kettle},
  \bibinfo{author}{J.~Koponen}, \bibinfo{author}{G.~P. Lepage},
  \bibinfo{author}{A.~T. Lytle}, \bibinfo{title}{{B-meson decay constants: a
  more complete picture from full lattice QCD}}, \bibinfo{journal}{Phys. Rev.
  D} \bibinfo{volume}{91} (\bibinfo{year}{2015}) \bibinfo{pages}{114509}.

\bibitem[{Lubicz et~al.(2017{\natexlab{b}})Lubicz, Melis, and
  Simula}]{Lubicz:2017asp}
\bibinfo{author}{V.~Lubicz}, \bibinfo{author}{A.~Melis},
  \bibinfo{author}{S.~Simula}, \bibinfo{title}{{Masses and decay constants of
  $D_{(s)}^\ast$ and $B_{(s)}^\ast$ mesons with N$_f = 2 + 1 + 1$ twisted mass
  fermions}}, \bibinfo{journal}{Phys. Rev. D} \bibinfo{volume}{96}
  (\bibinfo{year}{2017}{\natexlab{b}}) \bibinfo{pages}{034524}.

\bibitem[{Maris et~al.(1998)Maris, Roberts, and Tandy}]{Maris:1997hd}
\bibinfo{author}{P.~Maris}, \bibinfo{author}{C.~D. Roberts},
  \bibinfo{author}{P.~C. Tandy}, \bibinfo{title}{{Pion mass and decay
  constant}}, \bibinfo{journal}{Phys. Lett. B} \bibinfo{volume}{420}
  (\bibinfo{year}{1998}) \bibinfo{pages}{267--273}.

\end{thebibliography}

\end{document}